%% file: arxiv-OAMC-102523.tex
\documentclass{informs3}
\usepackage[]{caption}
\usepackage{subcaption}
\usepackage{float}
\usepackage{wrapfig}
\usepackage{booktabs}
\usepackage{amsmath}
\usepackage{amsfonts}
\usepackage{amssymb}
\usepackage{amsbsy}
\usepackage{bbm}
\usepackage{color, colortbl}
\usepackage{array}
\usepackage{mathtools}
\usepackage{multirow}
\usepackage{footnote}
\makesavenoteenv{tabular}
\usepackage{enumitem}
\usepackage{bm}
\usepackage[algo2e]{algorithm2e}
\usepackage{algorithm}
\DoubleSpacedXI

\usepackage{natbib}
 \bibpunct[, ]{(}{)}{,}{a}{}{,}%
 \def\bibfont{\small}%
\TheoremsNumberedThrough     % Preferred (Theorem 1, Lemma 1, Theorem 2)
\EquationsNumberedThrough    % Default: (1), (2), ...
\input{mycommands.tex} 
\begin{document}
%%%%%%%%%%%%%%%%

%\setcounter{page}{1}
%\DOI{10.1287/xxxx.0000.0000}%

\RUNAUTHOR{Vahdat and Shashaani}

\RUNTITLE{Robust Output Analysis}
%\RUNTITLE{}

% \TITLE{Bundling Information Goods of Decreasing Value}
\TITLE{Robust Output Analysis with Monte-Carlo Methodology}

\ARTICLEAUTHORS{%
\AUTHOR{Kimia Vahdat}
\AFF{ \EMAIL{kvahdat@ncsu.edu},\\ Edward P. Fitts Department of Industrial and Systems Engineering,\\ North Carolina State University \\
915 Partners Way, Raleigh, NC 27695}
\AUTHOR{Sara Shashaani}
\AFF{ \EMAIL{sshasha2@ncsu.edu},\\ Edward P. Fitts Department of Industrial and Systems Engineering,\\ North Carolina State University\\
915 Partners Way, Raleigh, NC 27695\\
\URL{https://shashaani.wordpress.ncsu.edu}}
} % end of the block

\ABSTRACT{ 
%\sara{update last} 
In predictive modeling with simulation or machine learning, it is critical to accurately assess the quality of estimated values through output analysis. In recent decades output analysis has become enriched with methods that quantify the impact of input data uncertainty in the model outputs to increase robustness. However, most developments are applicable assuming that the input data adheres to a parametric family of distributions. We propose a unified output analysis framework for simulation and machine learning outputs through the lens of Monte Carlo sampling. This framework provides nonparametric quantification of the variance and bias induced in the outputs with higher-order accuracy. Our new bias-corrected estimation from the model outputs leverages the extension of fast iterative bootstrap sampling and higher-order influence functions. For the scalability of the proposed estimation methods, we devise budget-optimal rules and leverage control variates for variance reduction. Our theoretical and numerical results demonstrate a clear advantage in building more robust confidence intervals from the model outputs with higher coverage probability.
%Although much progress has been made in quantification of In this paper, we propose a general framework facilitated with means to calculate its corresponding bias and variance for data-driven model performance analysis. Particularly, we focus on non-parametric methods to keep the proposed framework easily applicable to any problem. Iterative bootstrap sampling and higher-order influence functions are employed for bias and variance estimation. Numerical results over stochastic simulation and machine learning problem instances demonstrate the advantage of our method in building correct confidence intervals for the model's output.
}%

%\KEYWORDS{deterministic inventory theory; infinite linear programming duality; 
%  existence of optimal policies; semi-Markov decision process; cyclic schedule}

\HISTORY{}

\maketitle
%%%%%%%%%%%%%%%%%%%%%%%%%%%%%%%%%%%%%%%%%%%%%%%%%%%%%%%%%%%%%%%%%%%%%%

% Samples of sectioning (and labeling) in IJOC
% NOTE: (1) \section and \subsection do NOT end with a period
%       (2) \subsubsection and lower need end punctuation
%       (3) capitalization is as shown (title style).
%
%\section{Introduction.}\label{intro} %%1.
%\subsection{Duality and the Classical EOQ Problem.}\label{class-EOQ} %% 1.1.
%\subsection{Outline.}\label{outline1} %% 1.2.
%\subsubsection{Cyclic Schedules for the General Deterministic SMDP.}
%  \label{cyclic-schedules} %% 1.2.1
%\section{Problem Description.}\label{problemdescription} %% 2.

% Text of your paper here

\section{Introduction}
\label{sec:intro}
Estimating the output of a predictive logic has been studied for many decades in the statistics, simulation, and machine learning (ML) literature. Model output error estimation is used for two primary purposes \citep{raschka_model_2018}: (a) evaluating the performance of the predictive logic on unseen data (model generalization) and (b) adjusting for the settings of the predictive model and comparing different predictive modeling classes with each other. 

\begin{figure}[htp]
    \centering
    \includegraphics[width=8cm]{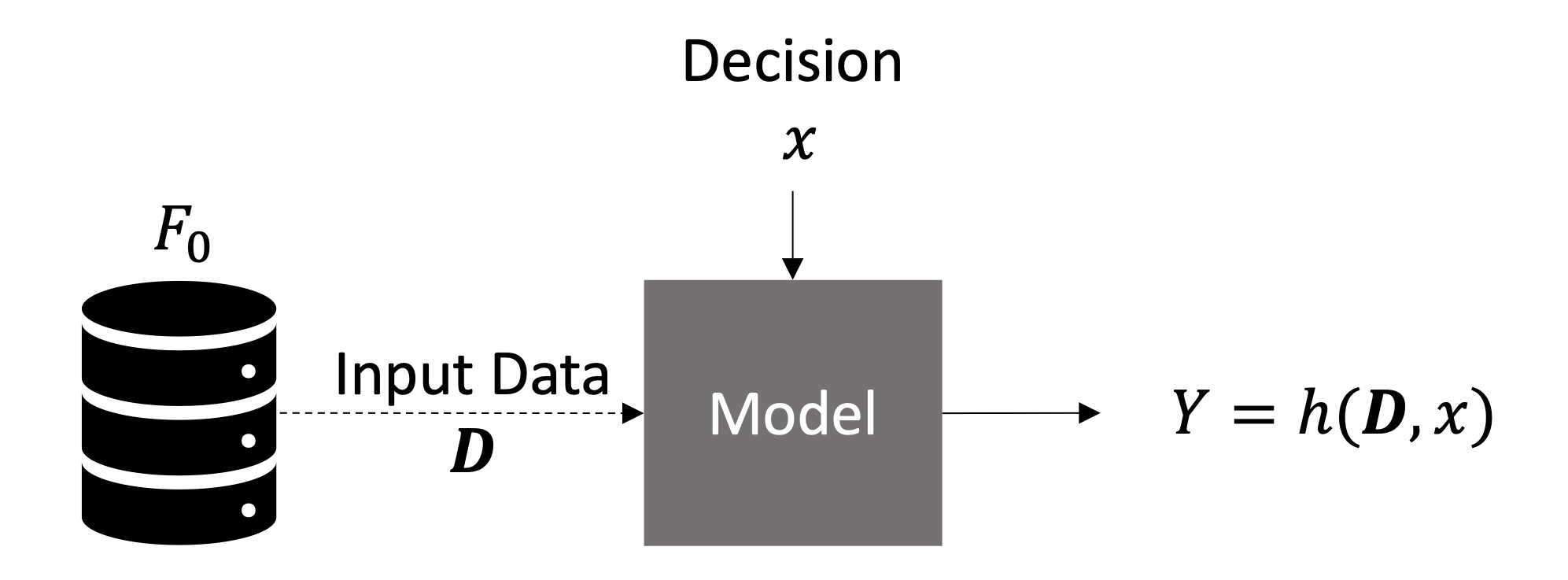}
    \caption{Diagram of model input-output relationship. Input data $\bm{D}$ is derived from a true unknown distribution, $F_0.$ The decision, $x,$ is entered to the model independently from the input data.}
    \label{fig:diag}
\end{figure}

Suppose that the model input denoted by $\bm{D}$, adheres to a true unknown distribution function $F_0$. The model output $Y=h(\bm{D},x)$ is an unknown functional of the model inputs and a control $x$ that is fixed and decided by the model user (see Figure~\ref{fig:diag}). Let the quantity of interest be the expected output of a model under the true unknown distribution of inputs that feed into the model, $\theta(F_0,x)=\mbE_{\bm{D}\sim F_0}[h(\bm{D},x)]$. Estimating $\theta(F_0,x)$ well means the ability to generate a confidence interval of it with an experiment that is guaranteed to cover $\theta(F_0,x)$ with high probability. Suppose $\thetahat(F_0,x)$ is the estimator along with the estimated $(1-\alpha)$\% confidence interval $\mcI_\alpha(x)$. Then (a) implies that $\Pr\{\mcI_\alpha(x)\ni \theta(F_0,x)\}\geq 1-\alpha$. Furthermore, (b) implies that if $x_1$ is a better choice for the model than $x_2$, i.e., $\theta(F_0,x_1)<\theta(F_0,x_2)$, then $\mcI_\alpha(x_1)$ lies below $\mcI_\alpha(x_2)$ with high probability. This is achieved by ensuring that the half-width of the two intervals is sufficiently narrow. 

In simulation, $h$ is a physics-based logic and $\bm{D}$ feeds into it as a random (set of) vector-valued variable(s). In ML, $h$ is data-driven and with a fixed logic, i.e., linear regression, $\bm{D}$ is the training set that produces the predicted outcome under decision $x$. For both, one may be interested in a decision $x$ that outperforms others. The model parameters that best relay the input-output relationship in simulation (validation/calibration) or level of complexity in ML (model selection) can also be viewed as decisions. In this case, $h$ represents some loss function between the model generated outputs and the real observed values and $\bm{D}$ is two (likely non-overlapping) sets for training and validation. 
% For both, $x$ could also be viewed as % one may be interested in finding the decision that outperforms others or 
% a set of parameters that best relay the input-output relationship -- validation or calibration in simulation -- or the model complexity -- model selection in ML. 

We remark that a typical approach to ML is to build one prediction with one training (and validation) set. In the Monte Carlo (MC) framework, this means the expected value is estimated with the output of one replication, which is a bad estimator. Instead, guided by the MC principle, multiple replications of training (and validation) sets provide multiple possible predictions (e.g., via multiple regression lines and associated coefficients).

In all of the above, there is a risk in using the models discussed that corresponds to our uncertainty about the true input distribution $F_0$. In fact, any error in the choice of $F_0$ leads to shifting (bias) and scaling (variance) in the model outputs' distribution. If both the bias and variance are quantified accurately enough, the goals in (a) and (b) can be achieved with more confidence. We term the added bias and variance to the model outputs as a result of uncertainty in the true input model $F_0$, the \emph{input uncertainty} (IU) induced bias and variance. 

While there is an extensive body of work for IU-induced variance and some recent attention to IU-reduced bias \citep{barton2022input,lam2016advanced,Song2019IU}%\sara{cite IU review papers here}
, the majority of these studies use parametric input models \citep{ChengHolland1997,morgan2019detecting} %\sara{cite parametric IU papers}
that may be most relevant to small input spaces with variables that are infrequently assumed to be dependent. However, the larger simulation models and data-driven applications consist of larger input domains with complicated dependency structures for which the parameteric input models would be restrictive and unjustifiable. Nonparametric approaches to input uncertainty, while not explicitly designed for simulation applications, were developed in the statistical learning community are through the bootstrap \citep{Hall2015double,LamQian2019,barton2018revisiting}. 

The drawback of the existing bootstrap approaches is their laborious computation in loops of resampled data for learning/fitting especially for a good estimation of the bias in the outputs. One level of bootstrapped samples of the data 
would achieve an estimate of the bootstrap standard deviation for an empirical estimator $\thetahat(x):=\theta(\Fhat,x)$, with $\Fhat$ denoting the empirical distribution of $\{D_i\}_{i=1}^n$. To achieve an estimate of the inflated variance in the outputs due to IU, using (i) new bootstraps from each bootstrapped resample (nested bootstrap) \citep{barton2018revisiting} and (ii) empirical delta method \citep{LamQian2019}. (i) requires resampling and repeated model running for a sufficient number of times to approximation the resample distribution. (ii) has a high variance even for parametric approximations and even more for the nonparametric regime. % with $\text{Var}(\thetahat(x))+R^{-1}\text{Var}(\thetahat(x))$. 
%Often the effect of small sample bias here is accounted for when computing the variance \cite{barton1993uniform}. 
Often the assumption of these studies, for $\Fhat$ considered as a random distribution for a random set of data $\{D_i\}_{i=1}^n$ is that the bias term in
\begin{equation}
    \theta(\Fhat)-\theta(F_0)\approx\mcN(\text{bias},\text{ IU variance}+\text{expected MC variance}
    % \text{Var}(\thetahat(x))+R^{-1}\text{Var}(\thetahat(x))
    )
    \label{eq:main-diff}
\end{equation} is 0 \citep{Song2019IU}. (in the remainder of the paper $\Fhat:=\Fhat|\{D_i\}_{i=1}^n$ is considered fixed.) Although this assumption holds as $n\to\infty$, with either small samples or changing input generating functions, the bias can be a significant factor in the estimator's error. In \eqref{eq:main-diff}, the variance term comes from a simple decomposition and use of total variance law. We have also removed $x$ from the notation for ease of exposition.

In the presence of an accurate bias estimator, one can correct the estimator with $\thetahat-\text{bias}$ before using it for inference or decision-making. However, compared to the variance, it is harder and more expensive to estimate the bias in either platform (simulation and ML). Parametric or nonparametric methods provide $\mcO(n^{-1})$ accuracy and often apply the bias directly to the confidence interval (by scaling/shifting the standard error or the critical value). Direct inference of the output bias or the bias of the estimator are not straightforward given the more realistic assumption of asymmetry in the outputs distribution. 
We explore estimating the bias using two non-parametric methods,(i) fast iterated bootstrapping (FIB), and (ii) higher order influence functions (HOIF). FIB \citep{ouysse2013,Hall2015double} utilizes nested iterative bootstrapping to estimate the bias to the order of $\mcO(n^{-3/2})$ with controlled computation cost. We explore augmenting FIB with simulation output analysis bridging the fields of simulation and ML. HOIF takes another approach toward estimating the bias; it estimates the second-order sensitivity of the output towards changes in the input distribution. 
Achieving $\mcO(n^{-3/2})$ order of accuracy in the bias, which is slightly better than the variance term but important for small samples, seems elusive.%\sara{check for correctness}\kimia{Looks Good}

% In this paper, our focus is on achieving (a) by ensuring that the true  \sout{ by correctly quantifying the effect of \emph{input uncertainty} (IU) on models' outputs}. Although goal (b) is not \emph{directly addressed}, methods developed for goal (a) are shown to be effective in achieving the latter goal (see Section~\ref{sec:ill2}). To motivate the benefit of our proposed robust output analysis, we provide two illustrative examples corresponding to (a) and (b) in sections~\ref{sec:ill} and \ref{sec:ill2}. 

\subsection{Contributions} 
We provide a framework for consistent and efficient quantification of IU-induced bias and variance with nonparametric input models. While our approaches are based on nested bootstrapping and nonparametric delta method, we devise \emph{efficient} ways by which higher order of accuracy for the bias estimates can be achieved. In particular, we provide the following contributions: 

% The practical stochastic model analysis platforms encompass ML and simulation. In ML, fitting a model to the observed data is challenging during model construction and hyperparameter tuning. The goal is for the model to capture the underlying patterns but not mimic the input data so closely that it fails to generalize. For simulation, resembling challenges are calibrating the logic model and analysis of output when there is uncertainty in input model estimation.%different scenarios for the system when there is limited observed data. \sara{reviewer says:  IU is not viewed as an approach to fix the issue that some system scenarios have limited observations}
% Simulation literature refers to the latter as IU with a track of parametric methods for modeling the input distributions for constructing inflated CIs on the estimated output. Computing CI requires an accurate estimation of model output bias and variance so that its actual coverage is close to its nominal value. 

% This issue is direr in the ML problems due to high-dimensional input datasets and the additional impact of IU on the model. Fitting a parametric distribution to high-dimensional data fails to capture the underlying uncertainty in the actual density generating function, $F_0$. Given the empirical distribution of the observed data, we develop a non-parametric estimator for model output, such that correct CI for \emph{any} model are easily attainable.  

\begin{itemize}
    % \item a unified framework for output analysis applicable to stochastic simulation and ML predictive modeling;
    \item combining fast iterated bootstrapping, non-parametric delta methods, and $m$-out-of-$n$ subsampling for asymptotically unbiased bias estimators; %non-parametric estimation methods for variance and bias of the model output utilizing the law of total variance, ;
    \item bias correction for each output rather than the overall estimator to leverage common random numbers and variance reduction through a novel nested control variate from simulation analysis; 
    \item minimizing the variance of the bias estimators using a multi-layer optimal budget allocation derived from nested simulation methodology; and
    \item proving the asymptotic validity of the resulting bias-corrected outputs' CI.
    % \item a procedure for out-of-bag sampling in computing the ML model prediction error incorporating the output bias and variance estimators.\sara{}% in addition to out-of-bag sampling  combined with $m$-out-of-$n$ bootstrapping;
    % \item demonstrating the benefits of employing the proposed approach on simulated datasets in both stochastic simulation and machine learning fields.
\end{itemize}

We allow the framework to be unified for output analysis applicable to stochastic simulation and ML predictive modeling. To that end, we redefine the notion of output analysis for ML by generating multiple outputs for a given $x$, using a model trained with an input set $\bm{D}$. This view towards ML, while reminiscent to cross-validation and other existing techniques, is formalized here with a MC lens that enables our proposed approaches to be applicable for data-driven settings. 
We also develop a procedure for out-of-bag sampling in computing the ML model prediction error that incorporates the output bias and variance estimator. %\sara{all the things unique to ML mentioned here.} 

During the development of our method, we became aware of a recently published non-parametric bias estimation method for stochastic optimization by \cite{iyengar2023bias}. They propose a bias estimator that utilizes a first-order influence function estimator. Their approach offers the advantage of not requiring additional model runs, resulting in improved efficiency.
In our research, we also achieved a similar outcome by utilizing a closed-form second-order influence function estimator, in addition to improving the order of accuracy of bias estimation.  
% \sara{isn't our work for higher order of accuracy in the bias than theirs?}\kimia{they also achieve a high order bias estimate.}\sara{really??}. 
By appropriately allocating our computational resources, we maintained the same total computing budget while providing an efficient bias estimator.

Ultimately, augmenting ML performance \emph{estimation} with Monte Carlo-based output analysis increases the reliability and robustness of ML's associated \emph{optimization} routines for model construction. Correctly estimating CI for the outputs gives us more accurate performance measures for a given input, which can tremendously help compare solutions and increase robustness. %\sara{this statement is not for here; see if you can use it in the intro}

% \sara{update at the end} 
In the following section, we first illustrate the proposed method's practicality concerning an inventory simulation problem. Section~\ref{sec:back} covers the background literature on both ML and simulation techniques for output analysis. Section~\ref{sec:method} defines the proposed estimator and demonstrates its statistical properties. Then, in Section~\ref{sec:ML}, we elaborate on the nuances of ML model prediction estimation. Section~\ref{sec:num} will compare the benefits of the proposed estimator with the existing benchmarks. Lastly, in Section~\ref{sec:conclusion}, we conclude our discussion and point the interested audience to future research directions.

\section{Illustrations -- An inventory simulation}
\label{sec:ill}
We now illustrate the impact of the proposed method on system performance estimate (goal (a)) and identification of the best alternative (goal (b)) in an inventory simulation example.
Take an $(s, S)$ inventory system for which demand data observations exist. Let $n$ be the number of demand observations on-hand. We replicate the inventory system used in \cite{KoenigLaw1985Sim} with some modifications. The inventory policy in this system is to order up to $S$ when the on-hand inventory falls below $s$. For this example, the underlying demand distribution per period follows a Poisson distribution, which is unknown during the simulation. We generate two demand datasets for this illustration. In the first example, the demand is generated from a Poisson distribution, where the average is 30, without any noise, and in the second example, the data generating function is a Poisson with an average of 25 and nonzero mean integer noise, which has an average of 5. In both examples, the expected demand rate is 30, and a goodness of fit test ensures both demand data significantly follow the Poisson distribution. Also, a large warm-up period (10,000) is set for both examples to ensure the observed demand and cost are stabilized. From this point on, we will refer to the first example as ``perfect Poisson'' and the second as ``corrupt Poisson''. The simulation logic is known, which generates the average total cost consisting of unit holding, ordering, and shortage costs over the simulation horizon (30 periods). In all the experiments, the total simulation budget (number of simulation runs) is fixed at 100, although the distribution of the budget differs across competing methods. 

\subsection{Goal (a): correct prediction}

For a given scenario, i.e., fixed $(s,S)=(20,45)$, suppose we have observed demand and the total cost for $n$ periods and aim to test the accuracy of the 30-period simulation output CI by its likelihood of covering the true expected total cost. This will help the stakeholder obtain a more reliable expectation of the magnitude of incurred costs and more reliable comparisons between alternatives.  %It will also be critical to comparing different decisions, i.e., goal (b) (see Section~\ref{sec:ill2} for a detailed example). 

%Denote the cdf of the true distribution, Poisson, for which the average demand rate is unknown, with $F_0$. 
Conventionally, assuming that the distribution family of the demand is known, methods such as maximum likelihood estimator (MLE) can estimate its parameter, say $\rhohat$ with the observed demands. Then, the fitted input model is incorporated into the simulation: the model produces replications of the output (expected total cost) under randomly generated inputs (demands) from the input model. We denote the simulation outputs with $Y_r(\Fhat):=h(\Fhat^{-1}(U_r),x)$, for $r=1,2,\ldots,R$ runs using $U_r\stackrel{iid}{\sim}\text{Unif}(0,1)$ and $\Fhat=F(\hat{\rho})$ to obtain $\bm{D}_r=\Fhat^{-1}(U_r).$
% \sara{in a parametric input model case this makes sense, but in a nonparametric case (empirical risk in ML; trace-driven simulation) where we ONLY use the data we have, isn't $r=1\cdots,n$ and then in subsampling $r=1\cdots,m$? If we made that change, I think it clarifies that we have $B_1$ outer distributions and $B_2$ inner distributions and an $n$ instead of $R$ will show up in our calculations}\kimia{I don't think so. $\bm{D}$ input in $h$ in ML cases, represents the training data or what $h$ is being built on. It needs to be a set of data points and not just one observation. Also, each $r$ simulation/ML model consist of a set of data points of size $n$ or $m.$ }
The function $h(\cdot,x)$ yields the average total cost after 30 periods for the ordering policy $x$. %\sara{notice how I have changed the notation you had $h(\bm{D}_r\sim\Fhat,x)$}. %where in this example $\Fhat=F(\hat{\rho})$.% such that $\hat{\rho}$ is the MLE (maximum likelihood estimator) of the input model's parameter. 
Then the desired expected total cost is $\theta(\Fhat,x)=\mbE_U[h(\Fhat^{-1}(U),x)]$ that can be estimated with $\thetahat(\Fhat,x)=\Ybar(\Fhat):=\frac{1}{R}\sum_{r=1}^R Y_r(\Fhat)$. We remark that using the data directly here and letting $\Fhat$ be the empirical distribution makes the sample average value yield $\theta(\Fhat,x)$ and not its estimate. %\sara{let me know if this is not clear}. %We refer to this output analysis as ``crude". 
In this ``crude" method, the CI of the expected value of interest, exploiting the central limit theorem (CLT), %\sara{no, this CI is bc of CLT; important for you to understand why}
is estimated by $\Ybar(\Fhat)\pm t_{R-1,1-\alpha/2}\sqrt{\sum_{r=1}^R(Y_r(\Fhat)-\Ybar(\Fhat))^2/(R-1)},$ where $t_{R-1,1-\alpha/2}$ is the student's t-distribution critical value at $1-\alpha/2$ quantile. %The crude method exploits the distributional properties of the input model to estimate the CI. %Explicitly, the MLE of the variance of the point estimator and the law of large numbers (LLN) help build the confidence intervals.

However, the crude output analysis cannot quantify the increase in the outputs variance due to the uncertainty in the input model. When that variance increase is correctly quantified, the coverage of the estimated CIs increases. As summarized in Section~\ref{sec:intro}, methods such as the nested simulation \citep{barton2018revisiting} that inflate the crude CI as a result of propagated input model error enhance the quality of the output analysis. %input data variance when the data is limited or noisy. 
%The crude method can be enhanced\sara{clarify what we mean by this; reviewer says:  The "crude" approach is not designed to output a confidence interval but instead a point estimator. It's a different goal from the authors so I'm not sure "enhanced" is the right word.} by inflating its variance with the IU-induced variance 
% The crude output analysis is not equipped for quantifying the input data variance, which is referred to as IU. The IU-inflated CI incorporates the variability of the unknown input distribution via a two-level sampling procedure \citep{barton2018revisiting}. 
This involves replacing $ Y(\Fhat)$ with $ Y(\Fhat^*)$ (in our example, $\Fhat^*=F(\hat{\rho}^*)$ --  Poisson distribution with rate $\hat{\rho}^*$ computed from resamples with replacement). Because there is no guarantee that the bias effect on the outputs is the same across all input models and simulation runs, our proposed approach tracks the debiasing of each output separately as $\widehat{\text{bias}}_r(\Fhat)$ use bootstrapped resample distributions $\Fhat^*$. Then we obtain $\thetahat(F_0)=\frac{1}{R}\sum_{r=1}^R (Y_r(\Fhat)-\widehat{\text{bias}}_r(\Fhat))$. 
\begin{remark}
    Although we use a parametric input model and use bootstrapping, our main difference with the existing parametric methods for bias estimation~\cite{kent2021bias,morgan2019detecting,morgan2022reducing,Song2019IU,Schuwirth2012bias} is that we do not use the difference between the input distributions with their parametric representations and instead compute those differences nonparametrically with empirical distributions. The proposed method will be described in Section~\ref{sec:method}.  This clarifies that although the proposed methods can still be used in conjunction with parametric input models, its generality extends to larger input spaces with complex codependence structures where the empirical distribution will approximate their joint probability distributions.%\sara{please confirm (check both Morgan's papers and cite others if any)}
\end{remark}

\begin{figure}%[H]
\centering
\includegraphics[width=\textwidth]{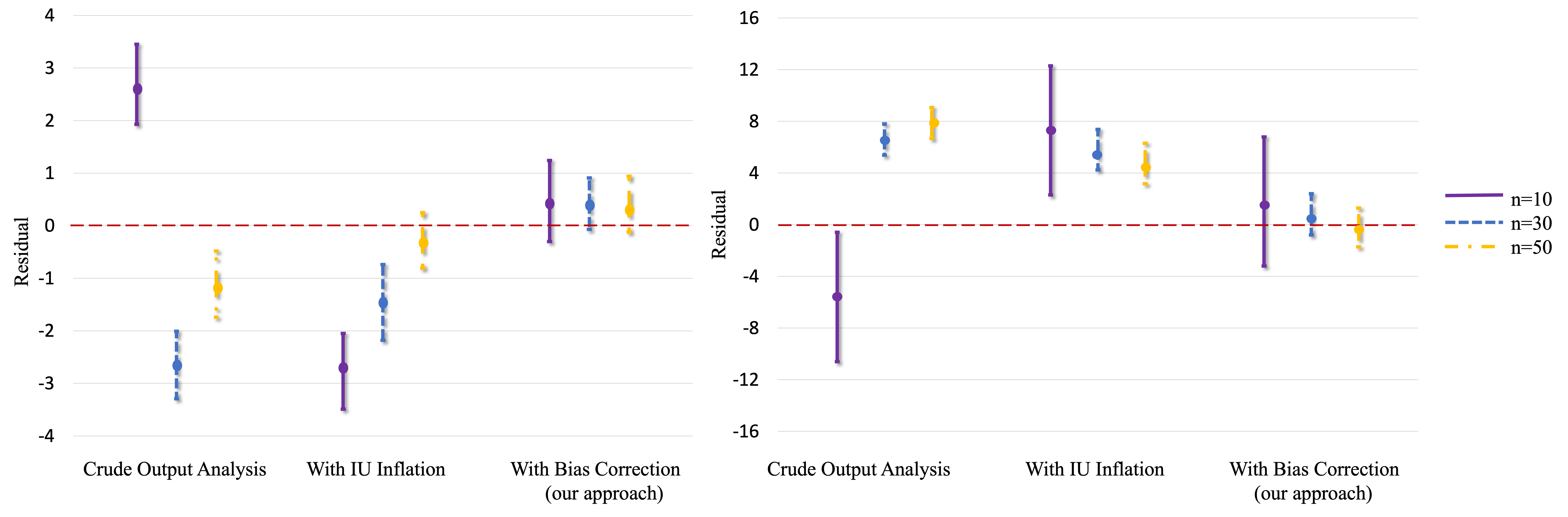}
%    \begin{subfigure}[b]{\textwidth}\centering
%    \includegraphics[width=0.75\textwidth]{./figs/Ill2CIPerfectV2.png}
%    \caption{}
%    \label{fig:illustration1} 
% \end{subfigure}\\
% \begin{subfigure}[b]{\textwidth}\centering
%    \includegraphics[width=0.75\textwidth]{./figs/Ill2CICorruptV2.png}
%    \caption{}
%    \label{fig:illustration2}
% \end{subfigure}
\caption{Three output analysis approaches are demonstrated over three input data sizes with $R=100$. The points represent the average difference between the estimated cost and the observed cost. The left panel shows the experiment with perfect Poisson. We observe that the CI are smaller than those in the corrupt Poisson case, in the right panel. Furthermore, the IU-inflated CI are able to encompass the true cost when $n$ is large in the perfect Poisson, which is not the case in the corrupt Poisson example. %\sara{when I download the figure, the x-axis labels appear strangely in a box; please fix. Add (our approach) under the bias correction term.} 
}
\label{fig:illustration}
\end{figure}

Figure~\ref{fig:illustration} demonstrates the effect of debiasing with the approach that we propose in this paper on the ability to predict in a range that covers the true average cost, $\frac{1}{30}\sum_{i=1}^{30} D_i$, where $D_i$ denotes the observed cost of the $i$-th period after the warm-up. Note that in both the simulation experiment and data generating process, we have set up a large warm-up period, 10,000 in these examples, to capture the long-run average cost. Interestingly, even if the true input distribution is not exactly Poisson, that is when we deliberately made an incorrect choice for the input model, the bias correction helps with recovering $\theta(F_0)$ with the CI that covers the (approximated) expected average cost. %,  for both perfect and corrupt Poisson cases. Including zero in the CI means we have captured the desired value. 
% There is a clear advantage in debiasing, as it helps covering the desired value in the CI. 
Repeating this experiment with several input sizes reveals that in the crude method, the performance does not necessarily improve with larger $n$ due to the discrepancy between the fitted and actual distributions. %There is also the effect of running the simulation for only $30$ periods while the expectation is to predict the long-run average cost.
%\sara{carify please: when $n=10$, do you use 10 values for demand and 10 for average cost but running the simulation for a very long time to produce costs that are closer to long-run value?  The way it is written this is not clear at all. If with $n=10$ your cost averages are also only for 10 periods, then we have a problem... b/c your truth has now changed to 10-period average not long-run average but you are running your simulations or 30 periods. We would want to keep the long-run average the ALWAYS be what we want to predict and then see how well we can do it when there is little data or lots of data} \kimia{I set a warm-up period of 10000 before computing the long run average over $n$ periods. The number of available demands is used for estimating $\rhohat$ which is then used in the simulation.} \sara{What I was trying to say was that as you change $n$, the size of your ``simulated" data should NOT change. In both cases your fitted Poisson should generate lots of data (some of it you discard for warm up) and use the same amount for estimating the outcome. Change in $n$ will only affect the discrepancy of the fitted Poisson; does this make sense?}\kimia{yes it does. In every simulation unit we are reporting the average of 30 periods not $n$ periods. I'm sorry I made a mistake in my previous answer. I misspoke here "over $n$ periods". Please take a look at the text, I made sure it is clear what is our simulation unit.} 
The improvement due to increased input data size is evident in the IU-inflated in the case of perfect Poisson, but it is slow in the corrupt Poisson case. On the other hand, the bias-corrected CIs succeed in both cases and with all choices of $n$ in this experiment.

\subsection{Goal (b): correct comparison}\label{sec:ill2}
%Take the same $(s,S)$ inventory system as an example and assume we have data observations for multiple scenario sets (see Table~\ref{tab:scenarios}).
%\subsection{Stochastic simulation use case}\label{sec:Simnum}
% This illustration aims to demonstrate corrected CI's impact on comparing different systems, which can be used for goal (b) in Section~\ref{sec:intro}. It should be noted that goal (b) remains inadequately tackled in the present illustration. Our intention is to offer a potential avenue for utilizing the suggested approach, without providing a comprehensive solution to achieving goal (b). To adequately address goal (b), a more detailed and extensive analysis would be necessary, but this is beyond the scope of the current study
We now compare our proposed debiasing method with the state-of-the-art bias and variance estimation methods in the simulation literature in terms of their ability to identify better alternatives. We demonstrate this by comparing the stationary trend between four $(s,S)$ scenarios (see Table~\ref{tab:scenarios}), which is computed via 100 batch means of 30-period ($s,S$) inventory simulation runs, against those trends from the output analysis with different techniques. The synthetically generated \emph{long-run} trend, generated by running the simulation model for 10,000 periods as a warm-up before collecting the target 30-period costs, with the batch means seeks to bury the dependence and nonstationarity in the outputs. Our goal is to recover the correct trend, which indicates that scenario 1 performs the best for both perfect and corrupt Poisson inputs. In the perfect Poisson use case, the order of preference is scenario 1, followed by scenarios 3, 4, and 2. Similarly, in the corrupt Poisson scenario, the order is scenarios 2, 3, and 4, following scenario 1. %...\sara{complete: e.g., scenario XX is the best, followed by ...}.% Table~\ref{tab:scenarios} shows four tested scenarios, with their true expected values using the actual input distribution and 100 batch means of 30-period ($s,S$) inventory simulation runs.

\begin{figure}
    \centering
    \includegraphics[width=\textwidth]{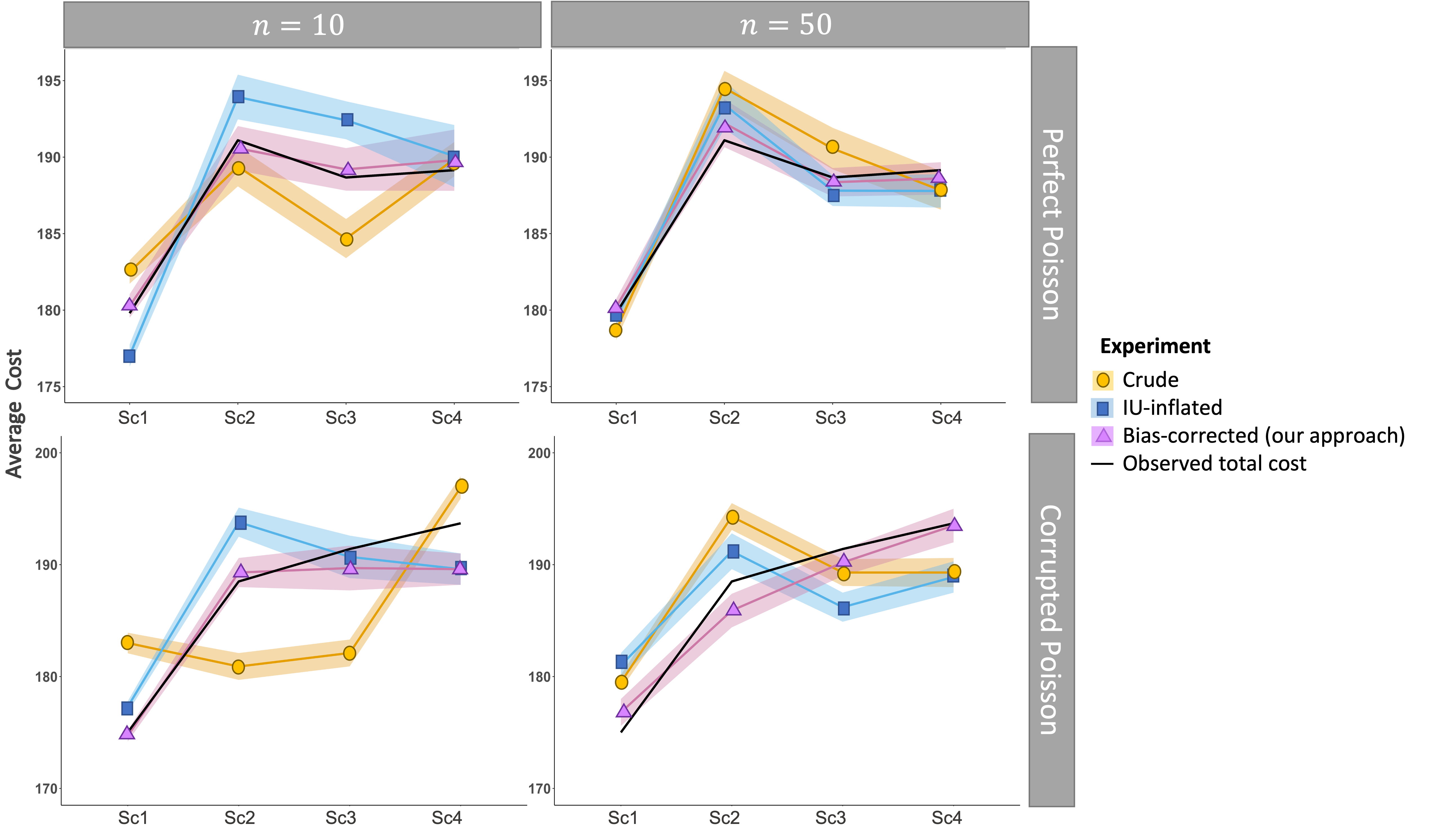}
    \caption{Comparing the 95\% CI for two input data sizes and three competing methods for the perfect and corrupt Poisson input datasets. The top row shows the true cost and CI for the perfect Poisson, and the bottom row shows the corrupted case. The observed total cost is different across the two distributions because it depends on the input data. %\sara{correct the y-axis label; also for x-axis don't use Sc in some places and sc in others please. Next to bias-corrected say (our approach)} 
    }
    \label{fig:comp}
\end{figure}

\begin{table}
    \TABLE
    {The four scenarios of the $(s,S)$ inventory problem used for illustration II. The last two rows denote the expected cost with perfect Poisson, $F^{\text{p}}_0$, and the corrupt Poisson, $F^{\text{c}}_0.$\label{tab:scenarios}}
    {\begin{tabular}{@{}lccccccc@{}}\toprule\centering

        Scenarios && 1 & 2 & 3 & 4  \\ \midrule
        $s$ && 20 & 20 & 20 & 20 \\
        $S$ && 40 & 45 & 50 & 55  \\
        $\theta(F^{\text{p}}_0)$ && 179 & 191 & 188 & 189  \\
        $\theta(F^{\text{c}}_0)$ && 175 & 186 & 191 & 195  \\
        %$\text{s.e.}(\oon\sumin \theta(M|D_i))$ && 0.11& 0.12 & 0.09 & 0.08  \\ 
        \bottomrule
        \end{tabular}}
    {}
\end{table}
The experiment is run for a fixed 30-period simulation using two input demand sizes of 10 and 50 along with the mentioned corrupt and perfect Poisson datasets. The input demand data is used for estimating input distribution and the simulation output reports the average total cost over 30 periods. Figure~\ref{fig:comp} reveals that the bias-corrected CI successfully captures the correct trend, and more resoundingly so with $n=50$. The input distribution greatly affects the crude method, as it does not allow any room for errors caused by IU. The distance between the crude CI and the observed cost roughly shows the magnitude of bias, which is in fact significant. This distance is more misleading in the corrupt Poisson dataset and $n=10$, as it suggests scenario 2 has the minimum average cost, whereas scenario 1 has the least expected cost. The IU-inflated method performs better than the crude method; however, it is far from identifying the correct order between scenarios. Furthermore, in the case of corrupt Poisson and $n=10$, the IU-inflated CI distorts the relationship among scenarios 2, 3, and 4 by displaying scenario 2 as having the highest cost and scenario 4 as having the lowest cost. However, this is the exact opposite of the actual relationship between the expected costs of these scenarios. Even when the data size increases to 50, although the crude and IU-inflated CI difference from the true expected cost appears to decrease, the true trend is still not accurately captured. % \sara{I could talk about the bottom left panel for example in the text to say where each method is failing; see my writing for the previous plot to improve the writing} 

% \begin{figure}
%     \FIGURE{\includegraphics[width=\textwidth]{./figs/Ill2CI.png}}{Comparing the 95\% CI for two input data sizes and three competing methods. %The black line shows the actual expected total cost. 
%     The role of bias is evident when comparing the different input data sizes. Although the bias decreases with larger input sizes, it plays an important role in correctly estimating the expected cost.\label{fig:comp}}{}
% \end{figure}

The bias-corrected method, particularly when $n$ is limited, accurately estimates the bias and variance, and successfully identifies the correct system, as well as the ranking of all systems. 
This observation promises that even training the models or validating them with bias-corrected outputs could yield better models to emulate the truth. In the simulation model example, the logic has a parameter on the number of periods to discard for warm-up. This parameter needs to be tuned. Instead of searching for the best $(s, S)$ values, our objective can be to tune this parameter with the bias-corrected outputs. We leave that experimentation for the interested reader. %Such a powerful output analysis can then be used to calibrate the simulation model with a tuned warm-up period parameter.  

Before describing the proposed approach for direct bias-correction, we first lay out a more extensive review of the literature for output analysis in both simulation and ML. 

\section{Background}
\label{sec:back}
Before describing the proposed approach for direct bias-correction, we first lay out a more extensive review of the literature for output analysis in both simulation and ML. 
% We now review the existing literature in simulation and ML output analysis. In simulation, $h$ is the model's underlying logic, often independent of data except in model validation. On the other hand, in ML, $h$ represents the data-dependent model's logic. In both $F_0$ is the input data distribution. 

\subsection{Stochastic Simulation}
% parametric IU
Output analysis has been widely studied under parametric and non-parametric distributions in simulation systems \citep{lam2016advanced, barton2012tutorial}. Parametric output analysis limits the data to a known family of distribution, which may fail to encompass the actual characteristics of the input data. Given a parametric input distribution, \cite{morgan2019detecting} compute an estimator for bias due to unknown input distribution for the simulation output. They achieve this by representing the simulation output as a function of the input distribution parameter. This allows them to rely on a vector of parameters to represent the input distribution. By employing Taylor expansion and central composite design around the MLE of the input parameter, they derive an estimator for the bias of the input distribution.
% \sara{you need to say more about their method since we are focusing on bias here; utilizing Taylor exansion doesn't really say what they do that is parametric} 
However, their method does not generalize to non-parametric input models.
\cite{Song2019IU} provide an extensive simulation output analysis under IU, where the bias and variance of the simulation models are studied considering parametric input models. They estimate the bias via a polynomial regression of degree 2 with data distribution parameters as independent variables. %\sara{correct?}\kimia{Yes}
% \sara{so they assume some form of a linear relationship between them or not?}\kimia{No, they do a polynomial regression with degree of 2, to find the bias.}
They show that the joint distribution between the input parameters and simulation output 
% \sara{what do you mean by distribution between the input and output? Do you mean joint distribution?} 
asymptotically follows a bi-variate normal, although with unclear validity for small data sizes ($n<50$).
\cite{kent2021bias} suggests another parametric approach towards deconvoluting the bias of IU in estimating a conditional expectation by debiasing the density of the input data 
% \sara{helps to use our notation to make the idea more clear; same for the ideas before this last one and those coming next; if you see the papers we have been using as samples, they all do this, explaining older methods in their new terminology to make the distinction between what exists and what will be new} \kimia{This one is hard to show with our notation. Because it uses the probability density function of the input data, different from cdf, and we need to introduce new notation to do so. I can work on it and find a way to summarize their approach after submitting the first draft of the thesis. }
Density deconvolution refers to fitting a density to noisy observed data by separating the noise from the density function.

% non-parametric IU
%To the best of our knowledge, without any prior distributional assumption on the input data, there is no procedure to identify the \emph{bias} due to input modeling. However, there are some noteworthy studies on estimating the variance of the simulation model output. 
Without any prior distributional assumption on the input data, there is one recent procedure to identify the \emph{bias} due to input modeling. \cite{iyengar2023bias} introduced a bias estimator that leverages the first-order functional derivation of the model output with respect to the empirical distribution, extending up to the order of $\mcO(n^{-1})$, to quantify the bias. They also require derivatives of the objective function, which in many practical cases is not available. This paper enhances the bias estimation with higher order IFs to $\mcO(n^{-2})$.
There are some noteworthy studies on estimating the variance of the simulation model output. \cite{davison1997App,barton1993uniform} and \cite{barton2001resample} present sampling method for variance estimation. They suggest a nested simulation framework with bootstrapping to capture the data variability. Later \cite{barton2018revisiting} present a variance reduction technique built on nested simulation framework to minimize the computation cost. 
%\sara{this is not the first time nested simulation was proposed; this study does something more, it uses shrinking to handle the cost by reducing the variance; please see Barton, Lam, and Song, 2022. Input uncertainty in stochastic simulation. In The Palgrave Handbook of Operations Research (pp. 573-620).}
Furthermore, \cite{LamQian2019} propose an alternative variance estimator using non-parametric delta methods, i.e., \textit{influence functions} (IFs). They estimate the variability of the output with respect to small changes in the empirical input distribution. Unlike nested bootstrapping, estimating IU variance with IFs does not require additional model runs, which makes it efficient, particularly when it is expensive to run a model. On the other hand, they require some known distributional properties, which can be replaced with empirical distribution known properties. %  \sara{maybe connect delta method with variance here and comment on the pros and cons of IF (this should motivate our use of IF in higher order)} 
%Lam and Qian's method can further be applied to estimating the gradient in optimization processes. 

% non-parametric estimation method in stat and ML
Non-parametric estimation methods are better suited for high-dimensional settings and are more generalizable to various problems. However, they have some drawbacks. Non-parametric methods, usually based on data sampling, are computationally expensive, so they may not perform well for a limited computation budget \citep{fithian2014optimal}. The majority of computational burden arises due to the requirement of performing nested simulations. Nested simulations are computationally demanding due to resampling at the outer layer and running simulation runs per resampled input model at the inner layer. To control the overall error affected by input and simulation variations, a large number of simulation runs are required, multiplied by the sampling effort in each layer \citep{barton2022input}. In this paper, we propose an optimal allocation of the computation budget for the proposed bias and variance estimators, inspired by budget allocation in nested simulation studies \citep{LamQian2021}, to efficiently encompass the underlying uncertainty of the observed data into our estimation. 
%\sara{you should connect this with budget allocation in nested simulation literature; also see section 17.3.2.2 of Barton, Lam, and Song, 2022} 

Another drawback of non-parametric methods in estimating the input distribution is their high dependence on the observed data \citep{lam2021impossibility}. We attempt to resolve this issue by implementing multi-level data sampling to reduce the dependency. More concretely, for $b_1=1,\cdots,B_1$ let the outputs of a simulation model with one level of sampling be, $\Ybar(\Fhat^*_{b_1})=\frac{1}{R}\sum_{r=1}^{R}Y_r(\Fhat^*_{b_1})$ and with two levels of sampling $\Ybbar(\Fhat^{**}_{b_1,.}),$ where %for an arbitrary $b_1\in\{1,\cdots,B_1\},$
$$\Ybbar(\Fhat^{**}_{b_1,.})=\frac{1}{R}\frac{1}{B_2}\sum_{r=1}^{R}\sum_{b_2=1}^{B_2}Y_r(\Fhat^{**}_{b_1,b_2}).$$ While both sets of outputs have a dependency on the observed input data, the dependency of the latter set is intuitively less than the former. The reason is $\Ybbar(\Fhat^{**}_{b_1,.})$s are averaged over $B_2$ perturbations of $\Fhat^*_{b_1}$ bootstrapped input model, mitigating the dependency on $\Fhat$.%the fixed input empirical distribution.
% \sara{this is not clear again; use notation to say what depends on what and why we think what we will do reduces dependency} \kimia{I think I made an error here. I don't think what we are doing actually addressing the dependence on observed data. Maybe I can say we are attempting to address this issue by conducting conditional analysis on the observed data? }\sara{The dependency between $\Ybar_{b}^*\ b=1\cdots B$ (if we didn't have a second layer) is less than $\bar\Ybar_{b}^*\ b=1\cdots B$ where $\bar\Ybar_{b}^*=R^{-1}\sum_{r}\Ybar_{b,r}^{**}$; right?}\kimia{I'm not sure. Isn't $\mbE[Y|\Fhat]=\mbE[\mbE[Y^*|\Fhat^*]|\Fhat]$?}

We have previously developed a bias estimator for the output of the stochastic simulation models using sampling-based methods and influence functions \citep{Vahdat2021}. Nevertheless, the estimated bias was not practical due to high variability. In this paper, we build on the previous bias estimator and reduce its variance by employing the variance reduction techniques. Additionally, we estimate the bias for each individual simulation output using higher orders of IFs (HOIF). The variance of these IFs estimators is optimized using variance reduction techniques and optimal budget allocation. Furthermore, we combine HOIF with our proposed reduced variance FIB to enhance the stability of the final bias estimator. This paper also demonstrates the asymptotic unbiasedness of the proposed estimator and establishes valid confidence intervals for the model's expected output. These confidence intervals are proven to asymptotically contain the true expected output with high probability. 
%rigorously prove the validity of the proposed estimator. %\sara{not just that! we also do things differently here in that we debias each output separately, use higher-order IF (could this help with lower variance?) and combine the two methods together (help with variance?)}. 
 %\sara{if you say this about our method, then you should also say it about the other methods you have named; which ones provide such a result? Can you be more specific about validity?}

% DRO
The benefits of model output analysis considering IU are not limited to evaluating the simulation models. An exciting application of IU that has been of focus recently is distributionally robust optimization (DRO). In DRO, minimizing the expected output of a model, typically considered in simulation optimization settings, $\min_x \theta(F_0,x)$, is converted to a min-max alternate, where the optimal solution is found within the worst-case in an uncertainty set, $\mcF$, i.e.,  $\min_{x}\sup_{F\in\mcF} \theta(F,x)$. 
The main challenge in DRO, similar to nonparametric delta method, is finding an unbiased and reliable estimate for $\nabla_{F}\theta(F,x)$. \cite{ghosh2018efficient} employ Giles' debiased estimator for the gradient in training ML models. 
%\sara{so Giles method is an alternative to IF?}\kimia{yes, only if there is no unbiased estimator of the desired function of $\theta(F,x)$. If we want to directly debias $\theta(F,x),$ Giles is no good.}. 
The Giles' method \citep{giles2008multilevel,blanchetGlynn2015} is a practical way of debiasing a function of expected values, $\theta(F,x)$. 
%\sara{where have we talked about expectation over $F$?}
However, it requires that the function does \textit{not} have an unbiased estimator, hence  it does not directly apply to sample average as an estimator of $\theta(F,x)$. \cite{lam2021distributionally} provide an unbiased estimator of stochastic gradient descent using only a few sample observations. Their proposed estimator utilizes score functions to cancel out the higher-order bias terms without explicitly characterizing the bias. It is important to note that their method is only helpful when we want to estimate difference in $x_1$ and $x_2$, which is the case in finite-differencing for gradient approximation; it does not apply to a general case of estimation for a given $x$. In this paper, however, we introduce a bias estimator directly applicable to $\theta(F,x)$ that can be directly used to guide the optimization. 

\subsection{Machine Learning}

The ML literature often does not recognize learning model outputs as a function of the \emph{unknown} input data distribution. The existing body of work contains methods for a limited set of learning models, i.e., linear regression. \cite{Shao1996Boot} and \cite{Rabbi2020LinRegBoot} introduce bootstrap-based confidence intervals for variable coefficients in linear regression that take the unknown input data into account. With the rise of black-box ML algorithms, there has been a new emphasis on model agnostic analysis, which, as its name suggests, is independent of the learning model \citep{efron2020Pred}. \textit{Model agnostic output analysis} relies primarily on ``good" data sampling procedures that provide a robust estimate and addresses the conditional nature of the model output. 
%\sara{the term ``model agnostic" suggests with any choice of learning alg. really; although the things you say here are nice and could be stated separately}. 
Good data sampling is critical, especially when the magnitude of the desired performance is important. 

$k-$fold cross-validation \citep{geisser1975cv,stone1974cross}, and several resampling methods such as $.632$ bootstrapping, double bootstrap, out-of-bag bootstrapping, and leave-one-out bootstrap or LOOBoot \citep{efron1997improvements,efron1983estimating} are well-known methods developed by statisticians and widely used for predictive model error analysis. $k-$fold cross-validation provides a balanced trade-off between bias and variance, with decreasing variance as the number of folds ($k$) increases. The bootstrapping method, which involves independently sampling training sets, yields low correlations between predictions and produces a small variance estimator. \cite{efron1983estimating}'s thorough analysis and numerical experiments reveal that both $k-$fold cross-validation and the bootstrap converge at a rate of $\mcO(n^{-2})$. However, the bootstrap estimator introduces additional terms that contribute to a downward bias. Techniques such as LOOBoot can partially mitigate this bias, aiming to enhance the accuracy of results at the cost of onerous computation. It is worth noting that none of these methods have a closed-form equation for bias estimation for the outputs, though there are some bias estimators for confidence intervals \citep{Hall1986}. %\sara{there are closed form bias estimations for the interval; see Hall 1986}. %\sara{talk about which one has high bias and why, and which high variance and why}. 
For a more extensive review of the bootstrap-based model output estimation techniques, we refer the interested readers to a survey by \cite{Austin2004bootstrap}. While these methods' estimates may be good enough for general model comparisons%\sara{so here we say getting good comparisons is easier than getting good predictions?}\kimia{yes}
, they fail to deliver a valid output CI. As noted, 
%\sara{this was not noted before or precisely called reliable CI; check Lam's papers for exact definitions. We have valid CIs, and asymptotically unbiased CIs, etc.} 
valid CI refers to confidence intervals that cover the true values with high likelihood, i.e., high coverage probability. We propose a novel bootstrap-based sampling method for quantifying the model's performance, resulting in an asymptotically unbiased point estimator expected model output with IU-inflated variance, thus a more reliable CI with improved coverage probability. 

Several studies in statistical analysis have introduced sampling-based estimators of bias. An interesting approach described by \cite{Hall2015double} and \cite{ouysse2013} is the double bootstrapping method, which involves perturbing the input data at two levels. The first level introduces perturbations to the data, while the second level varies the distributions obtained from the first level perturbations. This is directly linked to the use of the nested-simulation technique in stochastic simulation, which is employed to assess the variance of the input distribution. However, in multi-layer bootstrapping, the primary objective was to estimate the bias and enhancing the accuracy of the the output estimator. 
%\sara{this is almost exactly the same as the two level methods in simulation; draw connection and say what is different}
% \sara{my more important question is do any of these papers say to what order of accuracy they are able to estimate bias? Is it $\mcO(n^{-3/2})$? And are they debiasing the outputs directly or the interval?}\kimia{Their order of accuracy is $\mcO(n^{-3/2})$. They are debiasing the outputs directly.} 
Notably, the second level in multi-layer sampling techniques within statistical analysis only requires a single replication. Despite the potential of these estimators, they are often overlooked in stochastic simulation settings due to their computationally expensive nature. In our research, we propose a solution to address this limitation by introducing a fixed-budget optimal allocation alongside our bias estimator. This approach effectively alleviates the computational burden associated with previous methods. Moreover, our proposed budget allocation strategy aims to minimize the variance of the bias estimator, thus ensuring the reliability of the bias estimates. %\sara{why can't they be used for stoch sim? why is our method better than these? }. Also, optimizing the variability of their bias estimators and its impact on reliability is taken for granted \sara{not clear what you mean here}. 

\section{Proposed Methodology and Standing Assumptions}
\label{sec:method}
% We define our unified problem statement for ML and simulation as estimating $\theta(F_0,x)$ that measures the expected output of a logic model $h$ given the observed dataset, $\bm{D}$, that follows the unknown distribution $F,$ and input decision variable $x$. 
%We use $M(F_0)$ to recognize that the logic can be determined (ML) or calibrated (simulation) with the input model. 

%%%%%%%%% NEED TO MOVE THIS 
%Denote the on-hand dataset with $\bm{D}\in\mbR^{n\times (d+1)}:\{\langle \bm{D}_i,y_i\rangle\}_{i=1,\cdots,n}$, each point of which represents one input data point, $\bm{D}_i$, and its corresponding output, $y_i$.\sara{why do we need this here? can't it wait until the ML section?} 
%%%%%%%%%%%%%%%
%For ease of exposition, we simplify the model expected output with $\theta(F_0),$ as the decision is fixed for model output evaluation. 

Denote the on-hand dataset with $\bm{D}:\{ D_i\}_{i=1}^n$,%\sara{based on page 1, $\bm{D}$ is one input bc it gives one output. So in inventory model, it is one demand (or a vector of demands for 30 periods) for one replication. We instead of defining the empirical dist on its elements, I think we should define it as $\delta(\bm{D}_i)$. Right?},\kimia{$\bm{D}$ is one input, but it's a vector. Here, I want to refer to each element in this vector. The $\bm{D}$ is directly used as one input to train/build $h$, which then can create one output. Let me know if this is unclear. } 
each point of which represents one input data point. \emph{The objective is to estimate $\theta(F_0)$ accurately despite that $F_0$ is unknown.} %Let an output of the model $M(F_0)$ be represented with $Y(F_0)$. 
With $Y(F_0)$ as defined in Section~\ref{sec:intro}, we can write \[Y(F_0)=\theta(F_0)+\epsilon(F_0),\] where $\epsilon(F_0)$ is the MC error with mean 0~\citep{montgomery2009}. Given Assumptions~\ref{a:smooth} and ~\ref{a:simunbiased}, we aim to find an efficient and robust point estimator in addition to a valid CI for $\theta(F_0),$ where $F_0$ is not readily available. We call a CI valid, if its likelihood of containing $\theta(F_0)$ converges to 1, as the computation effort goes to infinity. In other words, with a valid CI, we are able to provide an estimate of the true expected model output, even when the input data is limited. 

\begin{assumption}\label{a:smooth}
    For an input model $F$ that is fixed over time, $\theta(\cdot)$ is a smooth function of $F$.
\end{assumption}
\begin{assumption}\label{a:simunbiased}
    Simulation outputs are conditionally unbiased, i.e., for any input model $F$ $\mbE[Y(F)|F]=\mbE_{U}[h(F^{-1}(U))|F]=\theta(F),$ where $U\sim Unif(0,1),$ is a uniform random number.
\end{assumption}

In practice $F_0$ can be estimated with the empirical distribution of the data on hand, $\Fhat=\oon \sumin \delta_{\bm{D}}(D_i)$, where $\delta_{\bm{D}}(D_i)$ denotes the Dirac measure (takes value of 1, if $D_i\in\bm{D}$, and 0 otherwise) for point $D_i$. 
% \sara{we have already talked about empirical distributions in section 1}\kimia{I think it helps with the flow if we briefly remind the reader where we start the process.} 
A point estimator for $\theta(\Fhat)$ is expressed as a sample average approximation (SAA) \citep{kleywegt2002sample} of $R$ simulation outputs, denoted by $Y_r(\Fhat)$ for $r=1,\cdots,R$. Here simulation refers to an iterative process of generating independent replications of $Y_r(\Fhat)$. Given Assumption~\ref{a:simunbiased}, the crude point estimator for $\theta(\Fhat)$ is\begin{align}
    \Ybar(\Fhat)=\frac{1}{R}\sum_{r=1}^R Y_r(\Fhat).\label{eq:estimator}
\end{align} Nevertheless, $\Fhat$ is merely one realization of $F_0,$ and $\Fhat\neq F_0$ and consequently, $\theta(\Fhat)\neq\theta(F_0).$ Of course, the empirical CDF converges in distribution to $F_0$ at an exponential rate \citep{massart1990empirical}, as $n$ tends to infinity. However, in many complex systems, either data is not readily available or using all of the data can be computationally expensive and we can opt for smaller subsets of data. Both cases results in $\beta(\Fhat):=\theta(\Fhat)-\theta(F_0)\neq 0,$ which induce an additional bias and variance into the estimation.

To capture the variability due to unknown input model, let $F$ be a random input model, then taking advantage of random effects model \citep{montgomery2009,ankenman2012quick} we expand the model output as a function of $F$, i.e.,
\begin{align}
    Y(F) = \theta(F_0)+\left(\theta(F)-\theta(F_0)\right)+\epsilon(F).\nonumber
\end{align}
\begin{align}
    Y_r(F)&=\theta(F)+\epsilon_r(F)\nonumber\\
    &=\theta(F_0)+\underbrace{\left(\theta(\Fhat)-\theta(F_0)\right)}_{\beta(\Fhat)}+\underbrace{\left(Y_r(F)-Y_r(\Fhat)\right)}_{W_r(F)}+\epsilon_r(\Fhat).\label{eq:Y-decompose}
\end{align} In (\ref{eq:Y-decompose}), we take advantage of the observed realization of $F_0,$  i.e., $\Fhat,$ to create an estimator for the model output. We define two sources of bias in (\ref{eq:Y-decompose}). $\beta(\Fhat)$ is the discrepancy between the expected output given the empirical distribution and true input model, and $W_r(F)$ is the random bias at each simulation output level. 
By the expansion in (\ref{eq:Y-decompose}), the MC error $\epsilon_r(F),$ is replaced with $\epsilon_r(\Fhat),$ which has less variance due to $\Fhat$ being fixed. Both bias terms can be negligible if the number of observed data points is large \citep{Hall1986}. 
% However, in many complex systems, either data is not readily available or using all of the data can be computationally expensive and we can opt for smaller subsets of data. 
We exploit two methods of higher order IFs and FIB to estimate $\beta(\Fhat),$ and $W_r(F),$ respectively.

\begin{remark}
    Alternatively, another expansion for $Y_r(F)$ is, \[Y_r(F)=\theta(F_0)+\left(Y_r(\Fhat)-Y_r(F_0)\right)+\left(Y_r(F)-Y_r(\Fhat)\right)+\epsilon_r(F_0),\] in which the MC error, $\epsilon_r(\Fhat),$ in (\ref{eq:Y-decompose}) is replaced with $\epsilon_r(F_0).$  There are two main issues with directly estimating $Y_r(\Fhat)-Y_r(F_0).$ The first being, the smoothness requirement of HOIF could not have been met with each output, whereas by LLN we could claim the smoothness for $\theta.$ Secondly, we found out that estimating $Y_r(\Fhat)-Y_r(F_0)$ would result in a highly variable bias estimator, which was not practical in ML examples.
\end{remark}   

%% Talk about estimating W and beta 
%% First introduce B_1 bootstrap distributions. 
We follow the bootstrap theory \citep{efron1979boot}, to imitate random input distributions from $\Fhat$. We denote $\Fhat^*_1:=\Fhat^*(\zeta_{1})$. We generate $B_1$ input models, $\Fhat^*_1,\cdots,\Fhat^*_{B_1}$,  %\sara{do not use $\Fhat^*(\zeta_{i})$ just like you are not using $h(\bm{D}_i)$ for $Y_i$} where $\zeta$ is a random uniform number indicating random perturbation in each bootstrapped dataset. 
for each of which we generate $R$ simulation outputs. Then (\ref{eq:Y-decompose}) can be written as
\begin{align}
    Y_r(\Fhat^*_{b_1}) = \theta(F_0) + \beta(\Fhat) + W_r(\Fhat^*_{b_1}) + \epsilon_r(\Fhat).\nonumber
\end{align} Section~\ref{sec:FIB} will elaborate on how to take advantage of the bootstrap theory, namely, $$\mbE_\zeta\left[\theta\left(\Fhat^*(\zeta)\right)\right]-\theta(\Fhat)=\theta(\Fhat)-\theta(F_0)+\mcO_p(n^{-1/2}),$$ with $\mbE_\zeta[\cdot]$ being the expectation with respect to the bootstraps perturbations, $\Fhat^*(\zeta),$ and fast iterated bootstrapping to find an efficient estimator for $W_r(\Fhat^*_{b_1})$ \citep{Hall2015double,ouysse2013}. The proposed estimator, $\What_r(\Fhat^*_{b_1})$ reduces the bias up to $\mcO(n^{-3/2}),$ which can significantly impact the decisions in smaller datasets.   
Also, the bias $\beta(\Fhat)$ cannot be directly observed because only the empirical distribution is known and is fixed for a given subset of data. Section~\ref{sec:biashoif} provides a closed-form estimator for $\beta(\Fhat),$ which we denote by $\betahat(\Fhat),$ that uses a functional expansion around the empirical distribution \citep{vaart1998}. Ultimately, estimating both bias terms, can improve our final estimator's accuracy to $\mcO(n^{-2}).$

% To create many random input datasets from $\Fhat$ and estimate the bias terms, we follow the bootstrap theory \citep{efron1979boot}, namely,  $$\mbE_\zeta\left[\theta\left(\Fhat^*(\zeta)\right)\right]-\theta(\Fhat)=\theta(\Fhat)-\theta(F_0)+\mcO_p(1/\sqrt{n}),$$ with $\mbE_\zeta[\cdot]$ being the expectation with respect to the bootstraps perturbations, $\Fhat^*(\zeta),$ where $\zeta$ is a random uniform number, to provide an approximation for $\beta(\Fhat)$. The same theorem can be stated, with some approximations, about each random output \citep{barton2018revisiting}, i.e., $$\mbE_\zeta\left[Y_r\left(\Fhat^*(\zeta)\right)\right]-Y(\Fhat)\approx Y(\Fhat)-Y(F_0),$$ which we leverage for estimating $W_r(\Fhat).$
% In other words, bootstrap theory enables estimating the bias, without accessing $F_0$, with unlimited computation power. The random bias in \eqref{eq:DefW} can be rewritten as
% \begin{equation}
%     W_r(\Fhat)\approx\mbE_\zeta\underbrace{\left[ \left. Y_r(\Fhat^*(\zeta))-Y_r(\Fhat)\right\vert\Fhat\right]}_{\Delta^*_r(\Fhat)}= \frac{1}{n^*} \sum_{b=1}^{n^*} \underbrace{Y_r(\Fhat^*_b)-Y_r(\Fhat)}_{\Delta^*_{r,b}(\Fhat)},\label{eq:biasFirst}
% \end{equation} where $n^*={{2n-1}\choose{n}}$ is the number of overall bootstrap possibilities with a dataset of size $n$, $\Fhat^*_b:=\Fhat^*(\zeta_b)$ is the $b$-th bootstrap's empirical distribution, and $\Delta^*_{r,b}(\Fhat)$ is the discrepancy between the random outputs with empirical and the $b$-th bootstrapped input distribution. 

%% Estimating variance
In addition to bias, estimating the total variance and each of its contributing components is critical in building a correct and valid CI and system diagnosis. When an estimate of each source of variability is known (stochastic noise or input data), one can effectively address the problematic component. For example, finding the input modeling process to contribute to the majority of variability can change the decision of conducting more simulation replications to collecting more data. % In this section, we estimate and decompose the total variance using the law of total variance. 
Define the debiased point estimator as,\begin{align}
    \Ybar^d(F) = \frac{1}{R}\sum_{r=1}^R Y_r(F)-\betahat(\Fhat)-\What_r(F), \label{eq:Y-decompose2}
\end{align} where $\mbE_U[\Ybar^d(F)]=\theta(F_0).$
Then its variance, given Assumption~\ref{a:U} can be written as,
\begin{align}
    \text{Var}_{U,F}(\Ybar^d(F))&= \text{Var}_{U,F}\left(\Ybar(F)-\frac{1}{R}\sum_{r=1}^R\What_r(F)\right)\nonumber\\
    &=\text{Var}\left(\Ybar(F)\right) + \text{Var}_{U,F}\left(\frac{1}{R}\sum_{r=1}^R\What_r(F)\right)-2\text{Cov}_{U,F}\left(\Ybar(F),\frac{1}{R}\sum_{r=1}^R\What_r(F)\right)\nonumber\\
    &= \frac{\sigma^2_{U}}{R}+\text{Var}_{F}(\theta(F)) + \text{Var}_{U,F}\left(\frac{1}{R}\sum_r\What_r(F)\right) - 2\frac{R-1}{R}\text{Cov}_{U,F}\left(Y_r(F),\What_r(F)\right),\label{eq:VarDecomposeDeb}
\end{align}where the first term quantifies the simulation variance, the second term is the IU variance, and the last two terms are variances associated with bias estimation. In Equation (\ref{eq:VarDecomposeDeb}), we expand $\text{Var}\left(\Ybar(F)\right)$ using the law of total variance, which involves conditioning on each source of uncertainty. For a detailed understanding of how this expansion works, refer to Remark 3.
% \begin{align}
%     \text{Var}(\Ybar^d(F))&=\mbE_{F}\left[\text{Var}_{U}\left(\Ybar^d(F)|F\right)\right]+\text{Var}_{F}\left(\mbE_{U}\left[\Ybar^d(F)|F\right]\right)\nonumber\\
%     &=\frac{1}{R}\mbE_F[\sigma^2_{U}(F)+\text{Var}_U(\What_r(F))-2\text{Cov}(Y_r(F),\What_r(F))]+\text{Var}_{F}(\theta(F))\nonumber\\
%     &=\frac{\sigma^2_{U}}{R}+\frac{\mbE_F[\text{Var}_U(\What_r(F))-2\text{Cov}_U(Y_r(F),\What_r(F))]}{R}+\text{Var}_{F}(\theta(F)),\label{eq:VarDecomposeDeb}\\
%     &= \frac{\sigma^2_{U}}{R}+\frac{\mbE_F[\text{Var}_U(\What_r(F))-2\text{Cov}_U(Y_r(F),\What_r(F))]}{R}+\text{Var}_{F}(\theta(F)+\mbE_U[\What_r(F)])\nonumber\\
%     &= \frac{\sigma^2_{U}}{R}+\frac{\mbE_F[-2\text{Cov}_U(Y_r(F),\What_r(F))]}{R}+\text{Var}_{F}(\theta(F))+2\text{Cov}_F(\theta(F),\What_r(F))\nonumber\\
%     &+\text{Var}(\mbE_U[\What_r(F)])+\mbE_F[\text{Var}_U(\What_r(F))]/R\nonumber\\
%     &= \frac{\sigma^2_{U}}{R}+\text{Var}_{F}(\theta(F))-2\frac{\mbE_F[\text{Cov}_U(Y_r(F),\What_r(F))]}{R}+2\text{Cov}_F(\mbE_U[Y_r(F)],\mbE_U[\What_r(F)])+\text{Var}(\frac{1}{R}\sum_r\What_r(F))\nonumber
%     %&=\frac{\mbE\left[\text{Var}(Y_r(\Fhat)|\Fhat)\right]}{R}+\text{Var}(\theta(\Fhat))\nonumber\\
% \end{align} 
% The $\text{Var}_{F}(\theta(F_0))$ in the second line of above expression is zero, as the inner term is not random. 
In Section~\ref{sec:var}, we further elaborate on how to estimate each component and how bias estimation affects the total variance.
\begin{assumption}\label{a:U}
    The variance of simulation output or stochastic uncertainty, i.e., $\sigma^2_{U},$ does not depend on its input model. Equivalently, $\mbE_F[\sigma^2_U(F)]=\sigma^2_U.$ 
\end{assumption}

\begin{remark}
    Without debiasing, the total variance of the biased estimator, under Assumption~\ref{a:U}, would be \begin{align}
    \text{Var}(\Ybar(F))&=\mbE_{F}\left[\text{Var}_{U}\left(\Ybar(F)|F\right)\right]+\text{Var}_{F}\left(\mbE_{U}\left[\Ybar(F)|F\right]\right)=\frac{\sigma^2_{U}}{R}+\text{Var}_{F}(\theta(F)),\label{eq:VarDecompose}
\end{align} where the first and second term is the same as stochastic variance in (\ref{eq:VarDecomposeDeb}). The difference between $\text{Var}(\Ybar(F))$ and (\ref{eq:VarDecomposeDeb}) captures the increased variability due to bias estimation.
\end{remark}

\begin{figure}
    \FIGURE{\includegraphics[width=\textwidth]{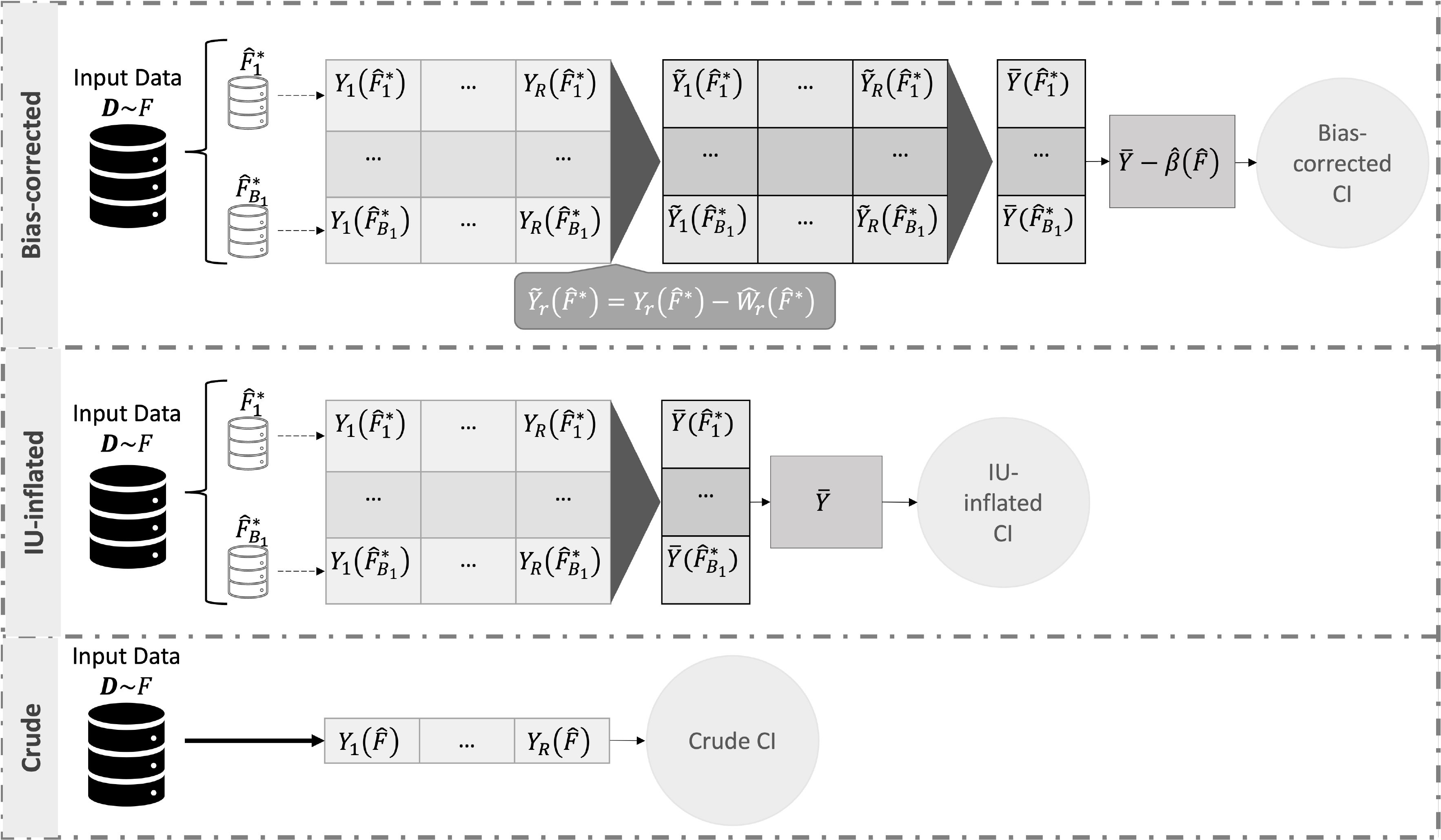}}{Steps to estimate the bias-corrected, IU-inflated, and crude CI are demonstrated. Figure~\ref{fig:biasEst} explains the algorithm for estimating $\betahat(\Fhat^*)$ and $\hat{\gamma}(\Fhat^*).$ The role of two bias estimates are illustrated in the top row. The second row includes the IU variance into the CI, and the last row is the crude CI.\label{fig:frame}}{}
\end{figure}

In the following subsections we detail the estimation methods for each term in (\ref{eq:Y-decompose}). 
From this point on, for ease of exposition, we replace $\mbE_{\zeta}[.]$ with $\mbE_{*}[.]$ and similarly, $\text{Var}_{\zeta}(.)$ with $\text{Var}_{*}(.).$

\begin{figure}
    \FIGURE{\includegraphics[width=\textwidth]{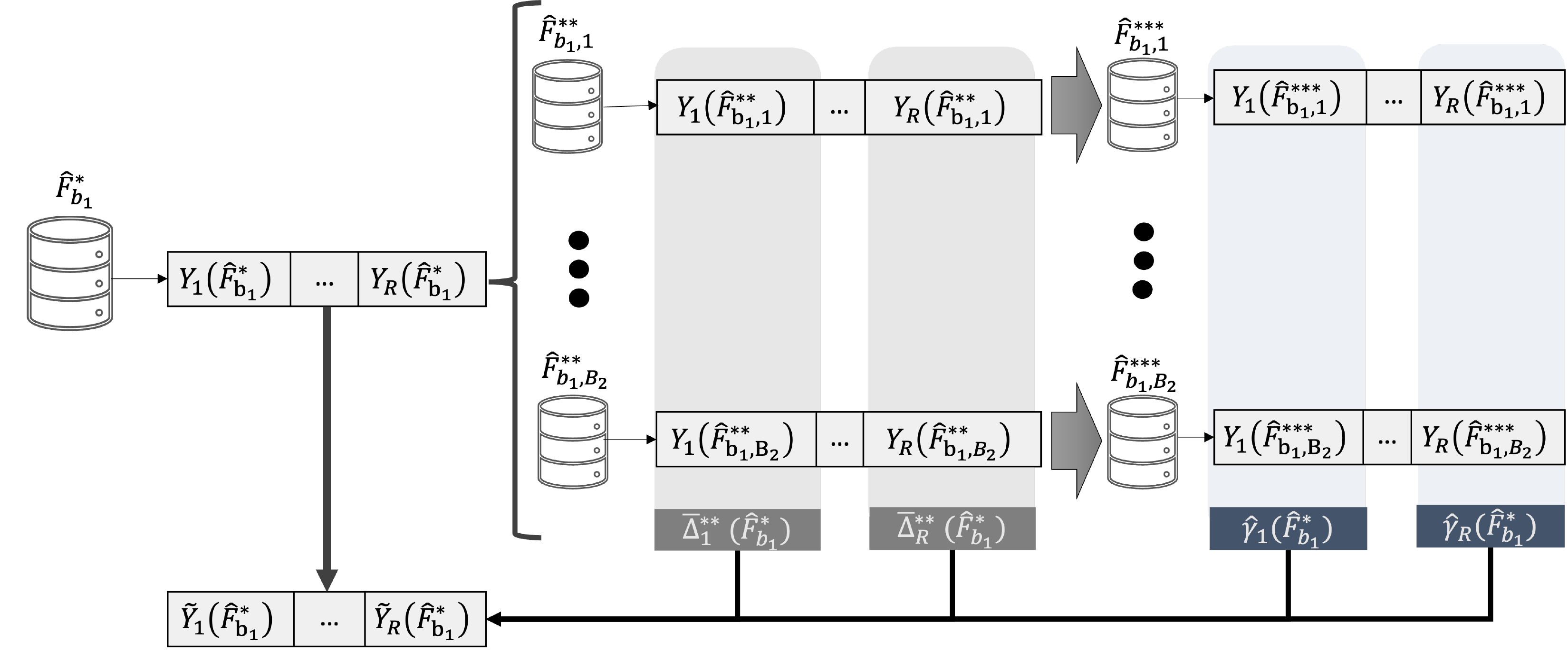}}{The proposed iterated bootstrap estimator is demonstrated.\label{fig:biasEst}}{}
\end{figure}

% \section{Estimation Procedures}\label{sec:estimation}

% This section elaborates on the methods used for estimating variance and bias that were briefly introduced in Section~\ref{sec:method}. Also, an optimized allocation of simulation budget is suggested. 

\subsection{Variance Decomposition}\label{sec:var}
This subsection provides an estimation for each term in (\ref{eq:VarDecomposeDeb}) using $B_1$ bootstrap resamples drawn for characterizing the input model.
Each $\Fhat^*_{b_1}$ for $b_1=1,\cdots,B_1$ is the input distribution of a sampled data $\{D_{b_1,i}\}_{i=1}^{n}$ drawn with replacement from $\Fhat$. The resampled datasets are conditionally independent of each other, and the simulation process is repeated $R$ times for each. Relying on Assumption~\ref{a:U}, estimating $\sigma^2_U,$ becomes a straightforward task using sum of squared errors of all simulated outputs ($RB_1$ total model outputs). We begin by estimating $\text{Var}_F(\theta(F)),$ then discuss an ANOVA approach for estimating the variance of bias estimators. 

With analysis of variance and bootstrap theory \citep{efron1979boot} (see Remark~3), 
\cite{barton2001resample,lam2016advanced} estimate the input distribution variance, i.e., $\text{Var}_F(\theta(F)),$ with $\text{Var}_*(\theta(\Fhat^*))$. Replacing random input models, $F,$ in \eqref{eq:VarDecompose} with bootstrapped distributions, $\Fhat^*,$ results in an estimator for IU variance as follows,
\begin{align}
    \widehat{\text{Var}}_*(\theta(\Fhat^*))&= \frac{1}{B_1-1}\sum_{b_1=1}^{B_1}\left(\Ybar(\Fhat^*_{b_1})-\Ybar\right)^2-\frac{1}{R}\frac{1}{B_1(R-1)}\sum_{b_1=1}^{B_1}\sum_{r=1}^R\left(Y_r(\Fhat^*_{b_1})-\Ybar(\Fhat^*_{b_1})\right)^2,\label{eq:VarANOVA}
\end{align} where $\Ybar(\Fhat^*_{b_1})=\sum_{r=1}^R Y_r(\Fhat^*_{b_1})/R$ and $\Ybar=\sum_{b_1=1}^{B_1}\Ybar(\Fhat^*_{b_1})/B_1$. In (\ref{eq:VarANOVA}) each variance is estimated with sum of squared residuals with respect to the variability source. We take advantage of their IU variance estimator directly in our total variance estimate.

Similar to the work of \cite{barton2001resample}, we calculate the sum of squared residuals of bias estimates and model outputs to find unbiased estimators for their variance and covariance terms. We conclude the total variance of the debiased estimator as, \begin{align}
    \widehat{\text{Var}}(\Ybar^d(\Fhat^*))&=\frac{1}{R(B_1R-1)}\sum_{b_1=1}^{B_1}\sum_{r=1}^{R}\left(Y_r(\Fhat^*_{b_1})-\Ybar\right)^2 \nonumber\\
    \ \ \ &+ \frac{1}{B_1R(B_1R-1)}\sum_{b_1=1}^{B_1}\sum_{r=1}^{R}\left(\What_r(\Fhat^*_{b_1})-\frac{1}{B_1}\sum_{b_1=1}^{B_1}\bar{W}(\Fhat^*_{b_1})\right)^2\nonumber\\
    \ \ \ &- \frac{2(R-1)}{B_1R^2(B_1R-1)}\sum_{b_1=1}^{B_1}\sum_{r=1}^{R}\left(\What_r(\Fhat^*_{b_1})-\frac{1}{B_1}\sum_{b_1=1}^{B_1}\bar{W}(\Fhat^*_{b_1})\right)\left(Y_r(\Fhat^*_{b_1})-\Ybar\right)\nonumber\\
    \ \ \ &+\frac{1}{B_1-1}\sum_{b_1=1}^{B_1}\left(\Ybar(\Fhat^*_{b_1})-\Ybar\right)^2-\frac{1}{R}\frac{1}{B_1(R-1)}\sum_{b_1=1}^{B_1}\sum_{r=1}^R\left(Y_r(\Fhat^*_{b_1})-\Ybar(\Fhat^*_{b_1})\right)^2,\label{eq:YDebTotVar}
\end{align}where $\bar{W}(\Fhat^*_{b_1}) = \frac{1}{R}\sum_{r=1}^{R}\What_r(\Fhat^*_{b_1}).$
Each estimated variance above identifies the contribution of its source (simulation, input data, or bias of input data) in the total variance and subsequently enhance the decision making. Next, we move onto estimating the biases introduced in Section~\ref{sec:method}

\subsection{Bias Estimation}
\label{sec:bias}
%Using the nested simulation as described in the previous section, we achieve a consistent estimator of the model error. 
%However, (\ref{eq:estimator1}) is biased 
Besides computing the variance of the output, we also need to quantify the bias, as described in (\ref{eq:Y-decompose}). Bias at the model output level can occur for two main reasons: the use of empirical distributions to quantify the error and the discrepancy between the true statistical and the estimated models. \emph{We only focus on the first bias term and leave the second one for future research.} 

In the previous section, in order to disintegrate the variance due to IU, we generated bootstrapped replications of the empirical distribution. We assume the expected model output is a smooth functional of input distribution and data observations are independent and identically distributed. Following the delta method a functional of the bootstrapped distributions converge to the functional of the actual distribution asymptotically, despite a non-negligible bias with limited simulation budget and data points \citep{vaart1998}. In the subsequent sections, we explore the two methods of FIB and HOIF for estimating the bias terms defined in (\ref{eq:Y-decompose2}). 

\subsubsection{Fast Iterated Bootstrapping}\label{sec:FIB}
We start by quantifying the bias due to bootstrapping, $W_r(\Fhat^*_{b_1})$. Each simulation output given the sampled input is denoted by $Y_r(\Fhat^*_{b_1})$ that is assumed to be a \emph{consistent} estimator of $Y(\Fhat)$. %Following the bootstrap theory \citep{efron1979boot} 
We write the bootstrap estimate of bias as, \begin{align}
    \bar{\Delta}_{r}^{*}(\Fhat^*_{b_1})
    &=\widehat{\mbE}_{**}[ Y_r(\Fhat^{*}_{b_1})-Y_r(\Fhat)]= \frac{1}{B_2}\sum_{b_2=1}^{B_2} Y_r\left(\Fhat_{b_1,b_2}^{**}\right)- Y_r(\Fhat^{*}_{b_1}),\label{eq:debiased1}
\end{align}where $B_2$ is the number of bootstrap resamples taken at each $B_1$ input models for identifying $W_r(\Fhat^*_{b_1})$. Also, $\Fhat^{**}_{b_1,b_2}$ represents the sample distribution taken with replacement from $\Fhat^*_{b_1}$. Note that in (\ref{eq:debiased1}) the expectation is taken over another level of bootstrapping and hence the bias estimator itself is biased with the order of $\mcO(n^{-3/2})$ \citep{Hall1986}. 

\cite{ouysse2013} shows that we can estimate the bias of (\ref{eq:debiased1}) with ``fast" bootstrap sampling, while maintaining asymptotic convergence of standard bootstrapping. Fast bootstrapping means that only one sample is taken for the second level to reduce the computation cost. Let $\gamma_r\left(\Fhat^{**}_{b_1,.}\right)$ be the residual bias of $\bar{\Delta}_{r}^{*}(\Fhat^*_{b_1})$ due to limited $B_2$, and $W_r(\Fhat^{*}_{b_1})=\bar{\Delta}_{r}^{*}(\Fhat^*_{b_1})+\gamma_r\left(\Fhat^{**}_{b_1,.}\right)$ 
be the total bias, then 
\begin{align}
    W_r(\Fhat^{*}_{b_1})=\mbE_{***}[ Y_r\left(\Fhat^{**}_{b_1,.}\right)-Y_r(\Fhat^{*}_{b_1})]=\bar{\Delta}^{*}_{r}\left(\Fhat^{**}_{b_1,.}\right)+\gamma_r\left(\Fhat^{**}_{b_1,.}\right),\label{eq:biastotal}
\end{align} where $\bar{\Delta}^{*}_{r}\left(\Fhat^{**}_{b_1,.}\right)=\widehat{\mbE}_{***}[ Y_r\left(\Fhat^{**}_{b_1,.}\right)- Y_r(\Fhat^{*}_{b_1})]$. Subtracting this equation from (\ref{eq:biastotal}) achieves, \begin{align}
    \hat{\gamma}_r\left(\Fhat^{**}_{b_1,.}\right)&=\widehat{\mbE}_{***}[ Y_r\left(\Fhat^{**}_{b_1,.}\right)-Y_r(\Fhat)]-\widehat{\mbE}_{***}[ Y_r\left(\Fhat^{**}_{b_1,.}\right)- Y_r(\Fhat^{*}_{b_1})]\nonumber\\
    &=\frac{1}{B_2}\sum_{b_2=1}^{B_2} \left( Y_r\left(\Fhat_{b_1,b_2}^{***}\right) - Y_r\left(\Fhat^{**}_{b_1,b_2}\right)\right) - \frac{1}{B_2}\sum_{b_2=1}^{B_2}  Y_r\left(\Fhat_{b_1,b_2}^{**}\right)+ Y_r(\Fhat^{*}_{b_1})\nonumber\\
    &=\frac{1}{B_2}\sum_{b_2=1}^{B_2}  Y_r\left(\Fhat_{b_1,b_2}^{***}\right) - \frac{2}{B_2}\sum_{b_2=1}^{B_2}  Y_r\left(\Fhat_{b_1,b_2}^{**}\right)+ Y_r(\Fhat^{*}_{b_1}).\nonumber
\end{align} 

We define the fast iterated bootstrap corrected estimator as\begin{align}
    Y^d_r(\Fhat^*_{b_1})+\betahat(\Fhat)&= Y_r(\Fhat^{*}_{b_1}) -\What_r(\Fhat^*_{b_1}) \nonumber\\
    &=Y_r(\Fhat^{*}_{b_1})-\bar{\Delta}^{*}_{r}\left(\Fhat^{**}_{b_1,.}\right)-\hat{\gamma}_r\left(\Fhat^{**}_{b_1,.}\right)\nonumber\\
    &=  Y_r(\Fhat^{*}_{b_1})+ \frac{1}{B_2}\sum_{b_2=1}^{B_2}  Y_r\left(\Fhat_{b_1,b_2}^{**}\right)-\frac{1}{B_2}\sum_{b_2=1}^{B_2}  Y_r\left(\Fhat_{b_1,b_2}^{***}\right). \label{eq:biasFIB}
\end{align}
The $\What_r(\Fhat^*_{b_1})=\bar{\Delta}^{*}_{r}\left(\Fhat^{**}_{b_1,.}\right)+\hat{\gamma}_r\left(\Fhat^{**}_{b_1,.}\right)$ is readily shown to be an unbiased estimator of the true bias of the model output given $\Fhat^*_{b_1}$, via the law of large numbers. \cite{ouysse2013} proves that fast bias approximation converges to the actual bias with $\mcO(1/B_{2}),$ and \cite{Hall1986} computes the error of the double iterated bootstrap to be of the order of $\mcO(n^{-3/2}).$ Lemma~\ref{thm:FIBunbiased} summarizes these properties.  
\begin{lemma}\label{thm:FIBunbiased}
   Assume $\mbE[\What_r(\Fhat^*_{b_1})^3]<\infty.$ By weak law of large numbers and bootstrap theory, 
   \begin{align}
        \big|\What_r(\Fhat^*_{b_1})-W_r(\Fhat^*_{b_1})\big|=\mcO_p(1/B_2).\nonumber
    \end{align}%where $\mcO_p$ denotes the order in probability.
\end{lemma}

\textbf{Significance Test for FIB Bias Estimator: }
To further enhance the robustness of the proposed estimator, we conduct a statistical significance test using the central limit theorem and asymptotic properties of bootstrapping. We introduce a pivotal statistic,\begin{align}
    T = \frac{\bar{\Delta}_{r}^{*}(\Fhat^{**}_{b_1,.})+\hat{\gamma}_r(\Fhat^{**}_{b_1,.})-W_r(\Fhat^*_{b_1})}{\sqrt{\frac{1}{B_2-1}\sum_{b_2=1}^{B_2}\left(\Delta^{*}_{r}(\Fhat^{**}_{b_1,b_2})+\hat{\gamma}_r(\Fhat^{**}_{b_1,b_2})-\What_{r}(\Fhat^*_{b_1})\right)^2}}, \label{eq:Tstat}
\end{align}which follows student's t distribution. 
\begin{theorem}\label{thm:convergance}
    Assuming standard regularity conditions, the pivotal statistic in (\ref{eq:Tstat}) follows an asymptotic student's t distribution. Subsequently\begin{align}
        \lim_{B_2\to\infty} \mbP\left\{ T_{\text{min}}(\alpha)\leq W_r(\Fhat^*_{b_1})\leq T_{\text{max}}(\alpha)\right\} = 1-\alpha,\nonumber
    \end{align}where \begin{align}
        T_{\text{max}}(\alpha) = \bar{\Delta}^{*}_r(\Fhat^{**}_{b_1,.})+\hat{\gamma}_r(\Fhat^{**}_{b_1,.})+t_{B_2-1,\alpha/2}\sqrt{\frac{1}{B_2-1}\sum_{b_2=1}^{B_2}\left(\Delta^{*}_{r}(\Fhat^{**}_{b_1,b_2})+\hat{\gamma}_r(\Fhat^{**}_{b_1,b_2})-\What_{r}(\Fhat^*_{b_1})\right)^2},\nonumber
    \end{align}and \begin{align}
        T_{\text{min}}(\alpha) =  \bar{\Delta}^{*}_r(\Fhat^{**}_{b_1,.})+\hat{\gamma}_r(\Fhat^{**}_{b_1,.})-t_{B_2-1,\alpha/2}\sqrt{\frac{1}{B_2-1}\sum_{b_2=1}^{B_2}\left(\Delta^{*}_{r}(\Fhat^{**}_{b_1,b_2})+\hat{\gamma}_r(\Fhat^{**}_{b_1,b_2})-\What_{r}(\Fhat^*_{b_1})\right)^2},\nonumber
    \end{align}where $t_{B_2-1,\alpha/2}$ is the $\alpha/2$ quantile of student's t distribution with $B_2-1$ degrees of freedom. 
\end{theorem}

\textbf{Reduced Variance FIB Bias Estimator: }
The proposed bias estimator takes advantage of two layers of resampling, which can increase the variance of the estimator. In this subsection, we explore a variance-reduced version of $\What_r(\Fhat^*_{b_1})$ using the control variate technique (\cite{glasserman2004monte}, chapter 4). The control variate is a variance reduction technique that adds a multiplier of a centered correlated variable, with a known expectation, to the estimator \citep{ross2022simulation}. The variance-reduced estimator improves the coverage probability by providing a less variable estimate of bias. We set the control variate as\begin{align}
    C_r(\Fhat^*_{b_1})= Y_{R+1}(\Fhat^{*}_{b_1})-Y_r(\Fhat^*_{b_1}),\nonumber
\end{align}where $Y_{R+1}(\Fhat^{*}_{b_1})$ is a model output generated from a seed that is independent from random seeds generating outputs $Y_1(\Fhat^*_1),\cdots,Y_R(\Fhat^*_{b_1}).$ Note that, $\mbE[C_r(\Fhat^*_{b_1})]=0$ as both terms are outputs of the same simulation model. 

Hence, the variance-reduced bias estimator will be (for a given $b_1$ and $r$)\begin{align}
    \What^{\text{cv}}_r(\Fhat^*_{b_1},c_1)=\What_r(\Fhat^*_{b_1})+c_1 C_r(\Fhat^*_{b_1}),\label{eq:Wreduced}
\end{align} where $c_1$ is a constant multiplier called coefficient of variation.
The optimized coefficient of variation can be found by minimizing the variance of $\What^{\text{cv}}_r(\Fhat^*_{b_1})$ with respect to $c_1.$ With some simple calculation we can see that the desired minimizer is $c_1^*=\text{Cov}(\What_r(\Fhat^*_{b_1}),C_r(\Fhat^*_{b_1}))/\text{Var}(C_r(\Fhat^*_{b_1})).$ Note that, $c_1^*$ is not directly attainable, but can be estimated via \begin{align}
    \chat_1^* = \frac{\sum_{r=1}^R\left(\What_r(\Fhat^*_{b_1})-\sum_{r=1}^R\What_r(\Fhat^*_{b_1})/R\right)\left(C_r(\Fhat^*_{b_1})-\bar{C}(\Fhat^*_{b_1})\right)}{\sum_{r=1}^R\left(C_r(\Fhat^*_{b_1})-\bar{C}(\Fhat^*_{b_1})\right)^2},\nonumber
\end{align}where $\bar{C}(\Fhat^*_{b_1})=\sum_{r=1}^RC_r(\Fhat^*_{b_1})/R.$ As $R$ tends to infinity, by the law of large numbers, $\chat_1^*$ converges to $c^*_1$ with probability 1. Nevertheless, substituting $c_1^*$ with its estimator induces a bias in the order of $\mcO(1/R)$ to $\What^{\text{cv}}_r(\Fhat^*_{b_1},c_1^*)$ \citep{glasserman2004monte, Nelson1990CV}. Lemma~\ref{thm:FIBCVunbiased} concludes that the bias of the variance-reduced FIB bias estimator converges to 0, as the computational effort, i.e., $B_2$ and $R$, grows large.
\begin{lemma}\label{thm:FIBCVunbiased}
    By Lemma~\ref{thm:FIBunbiased} and triangle inequality, we have
   \begin{align}
        \mbE\left[\big|\What^{\text{cv}}_r(\Fhat^*_{b_1},\chat^*_1)-W_r(\Fhat^*_{b_1})\big|\right]&=\mbE\left[\big|\What_r(\Fhat^*_{b_1})+\chat^*_1 C(\Fhat^*_{b_1})-W_r(\Fhat^*_{b_1})\big|\right]\nonumber\\
        &\leq \mbE\left[\big|\What_r(\Fhat^*_{b_1})-W_r(\Fhat^*_{b_1})\big|\right]+\mbE\left[\big|\chat^*_1 C(\Fhat^*_{b_1})\big|\right]\nonumber\\
        &= \mcO(\frac{1}{B_2})+\mcO(\frac{1}{R}).\nonumber
    \end{align} 
\end{lemma}
The variance-reduced final estimator of the model output (see (\ref{eq:biasFIB})) becomes \begin{align}
    \Ybar^{d,\text{cv}}_r(\Fhat^*_{b_1})+\betahat(\Fhat) &= Y_r(\Fhat^*_{b_1})-\What^{\text{cv}}_r(\Fhat^*_{b_1},\chat^*_1)\nonumber\\
    &= Y_r(\Fhat^{*}_{b_1})+ \frac{1}{B_2}\sum_{b_2=1}^{B_2}  Y_r\left(\Fhat_{b_1,b_2}^{**}\right)-\frac{1}{B_2}\sum_{b_2=1}^{B_2}  Y_r\left(\Fhat_{b_1,b_2}^{***}\right)-\chat^*_1(Y_{R+1}(\Fhat^*_{b_1})-Y_r(\Fhat^{*}_{b_1})). \label{eq:biasCVFIB}
\end{align}
\subsubsection{Higher Order Influence Functions}
\label{sec:biashoif}
%To quantify $\mbE[\theta(\Fhat^*)-\theta(\Fhat)]$, we begin with quantifying 

In this subsection, we quantify $\beta(\Fhat)=\theta(\Fhat)-\theta(F_0)$ using Von-Mises expansion and influence functions \citep{vaart1998}.
Von-Mises expansion is similar to the Taylor expansion with some modifications; define the function $\phi: t\to \theta(F+t(\Fhat-F_0)\sqrt{n})$. We estimate $\phi(t)$ at $t=0$ using the Taylor expansion as,
\begin{align}
    \theta(F+t(\Fhat-F_0)\sqrt{n})=\theta(F_0)+ t\nabla_{F}\theta(\Fhat-F_0)\sqrt{n}+\frac{nt^2}{2}\nabla^2_{F}\theta(\Fhat-F_0)^2 +\mcO((t\Vert\Fhat-F_0\Vert\sqrt{n})^3),\nonumber
\end{align} where $\nabla_{F}\theta$ and $\nabla^2_{F}\theta$ are the first and second order directional derivatives of $\theta$. In Von-Mises expansion, $t$ is replaced with $1/\sqrt{n}$ which results in, \begin{align}
    \theta(\Fhat)=\theta(F_0)+ \nabla_{F}\theta(\Fhat-F_0)+\frac{1}{2}\nabla^2_{F}\theta(\Fhat-F_0)^2 +\mcO(\Vert\Fhat-F_0\Vert^3).\label{eq:VonMises}
\end{align} 
Assuming a linear and continuous first-order derivative, we have $\mbE[\nabla_{F}\theta(\Fhat-F_0)]=\nabla_{F}\theta(\mbE_F[\Fhat-F])=0$. Hence, taking an expectation with respect to $F_0$ from (\ref{eq:VonMises}) yields $\mbE[\theta(\Fhat)-\theta(F_0)]\approx \mbE[\nabla^2_{F}\theta(\Fhat-F_0)^2]/2.$

Provided that the desired functional is smooth, i.e. infinitely many differentiable, \cite{Efron2014estaccuracy} shows that based on the bootstrap theory $\Vert \Fhat^*-\Fhat\Vert\to \Vert \Fhat-F_0\Vert$ as the number of data points grow large, and similarly, $\theta(\Fhat^*)-\theta(\Fhat)\to\theta(\Fhat)-\theta(F_0)$. The smoothness assumption is valid in our case because $\theta$ is the expectation of model error. Employing the smoothness of $\theta$, we can estimate its first and second-order derivatives using the bootstraps already created for variance estimation. Consequently, we rewrite (\ref{eq:VonMises}) using the empirical distribution and its random perturbation, $\Fhat^*$. 

Each $\Fhat^*_{b_1}$ for $b_1=1,\cdots,B_1$, as explained in Section~\ref{sec:var}, is sampled from the empirical distribution of the on-hand dataset, $\Fhat$, with replacement and size $m\leq n$. As noted by \cite{LamQian2021}, sampling less than the on-hand data size will help manage the computation cost while controlling the variance. We point out additional advantages of sub-sampling in ML application in Section~\ref{sec:ML}.

Define the probability of selecting point $i$ in $\Fhat^*_{b_1}$ as $N_{b_1,i}/n$, where \begin{align}
    N_{b_1,i}=\#\{\bm{D}_{b_1}=D_i\}\sim \text{binomial}(m,\bm{p}_0)\label{eq:NMult}
\end{align} is the number of repeated samples of $D_i$ in $\bm{D}_{b_1}$. 
In (\ref{eq:NMult}), $\bm{p}_0=(1/n,\cdots,1/n)$, and $\bm{N}_{b_1}\sim $Mult$(m,\bm{p}_0)$. Additionally, $\bm{D}_{b_1}$ represents the $b_1$-th sample taken from $\bm{D}$. 
Note that $\Fhat$ and $\Fhat^*$ can be written as $\frac{1}{n}\sum_{i=1}^n\delta(D_i)$ and $\frac{1}{m}\sumin N_{b_1,i}$, respectively. % where $N_{b_1,i}$ follows a Multi-nomial distribution that indicates the count of point $i$ in a selected set of $m$ points out of $n$ points. 
Then the Von-Mises expansion for an arbitrary $b_1$ is, \begin{align}
    \theta\left(\Fhat^*_{b_1}\right)&=\theta(\Fhat)+\sum_{i=1}^n\nabla_{\Fhat}\theta\left(\frac{N_{b_1,i}}{m}-\frac{\delta(D_i)}{n}\right)+\frac{1}{2}\sum_{i=1}^n\sum_{j=1}^n\nabla^2_{\Fhat}\theta\left(\frac{N_{b_1,i}}{m}-\frac{\delta(D_i)}{n}\right)\left(\frac{N_{b_1,j}}{m}-\frac{\delta(D_j)}{n}\right)\nonumber\\
    &=\theta(\Fhat)+\sum_{i=1}^n\nabla_{\Fhat}\theta\left(\frac{N_{b_1,i}}{m}-\frac{1}{n}\right)+\frac{1}{2}\sum_{i=1}^n\sum_{j=1}^n\nabla^2_{\Fhat}\theta\left(\frac{N_{b_1,i}}{m}-\frac{1}{n}\right)\left(\frac{N_{b_1,j}}{m}-\frac{1}{n}\right),\nonumber
\end{align} where the second equation simplifies the first by limiting the summation to the available data points and replacing $\delta$ with 1.

It remains to propose an unbiased estimator for $\nabla_{\Fhat}\theta$ and $\nabla^2_{\Fhat}\theta$. We employ score functions to develop the desired estimators. \cite{LamQian2019} propose an unbiased estimator for $\nabla_{\Fhat}\theta$ at point $D_i$ using score functions,\begin{align}
    \widehat{\text{IF}}_1\left(D_i;\Fhat\right)&=\frac{1}{B_1}\sum_{b_1=1}^{B_1}\frac{1}{R}\sum_{r=1}^{R} Y^d_r\left( \Fhat^*_{b_1}\right)S^{(1)}_i\left(\Fhat^*_{b_1}\right),\label{eq:IF1}
\end{align} where \begin{align}
    S^{(1)}_i\left(\Fhat^*_{b_1}\right)=\frac{n-1}{n\text{Var}(N_{b_1,i}/m)}\left(\frac{N_{b_1,i}}{m}-\oon\right)=mn\left(\frac{N_{b_1,i}}{m}-\oon\right)\nonumber
\end{align} is the score function. They show that $\mbE[\widehat{\text{IF}}_1\left(D_i;\Fhat\right)|\Fhat]=\nabla_{\Fhat}\theta$, hence it is unbiased. We build on their approach to provide an unbiased estimator for $\nabla^2_{\Fhat}\theta$, which then can be used for the bias estimation. 

Note that $\nabla^2_{\Fhat}\theta$ is a bilinear mapping, so for a given pair of distinct points $D_i$ and $D_j$, we define the estimator as,\begin{align}
    \widehat{\text{IF}}_2\left(D_i,D_j;\Fhat\right)&=\frac{1}{B_1}\sum_{b_1=1}^{B_1}\frac{1}{R}\sum_{r=1}^{R} Y^d_r\left( \Fhat^*_{b_1}\right)S^{(2)}_{i,j}\left(\Fhat^*_{b_1}\right)+\frac{\lambda Y^d_r(\Fhat)}{mn^2}-\lambda\eta\widehat{\text{IF}}_1\left(D_i;\Fhat\right),\label{eq:IF2}
\end{align} where \begin{align}
    S_{i,j}^{(2)}\left(\Fhat^*_{b_1}\right)&=\lambda\left(\frac{N_{b_1,i}}{m}-\oon\right)\left(\frac{N_{b_1,j}}{m}-\oon\right),\nonumber\\
    \lambda &\approx -1/5\text{Cov}\left(N_{b_1,i}/m,N_{b_1,j}/m\right),\label{eq:lambda}
\end{align} and \begin{align}
    \eta\approx 6\text{Cov}(N_{b_1,i}/m,N_{b_1,j}/m)+4\mbE[N_{b_1,i}/m]^3. \label{eq:eta}
\end{align} Appendix~\ref{sec:proofs} elaborates on derivation of (\ref{eq:lambda}) and (\ref{eq:eta}). Note that calculating $S_{i,j}^{(2)}\left(\Fhat^*_{b_1}\right)$, for all pairs of $(i,j)$, requires $n^2$ arithmetic computation, which grows large as the data size increases. This computation burden should be negligible in practice, since the proposed method's main application is for smaller datasets, in which input data induced bias is significant.
% \begin{align}
%     \frac{2}{\lambda}&=\frac{m(m-1)(m-2)(m-3)}{m^4n^2}+\frac{m(m-1)(m-2)}{m^3n^3}\left(\frac{5n}{m}-4n\right)\nonumber\\
%     &\ \ \ +\frac{m(m-1)}{n^2m^2}\left(\frac{4}{m^2}+\frac{8}{mn}-\frac{8}{mn^2}+6\right)-\frac{4}{mn^3}-\frac{3}{n^2}-\frac{2}{m^3n}+\frac{5}{mn^2},\nonumber\\
%     &=\frac{10}{mn^2}-\frac{8}{m^2n^2}+\frac{4}{mn^3}-\frac{8}{mn^4}-\frac{8}{m^2n^3}+\frac{8}{m^2n^4}-\frac{2}{m^3n}=\frac{10}{mn^2}+\mcO\left(n^{-4}\right).\label{eq:lambda}
% \end{align} Hence $\lambda\approx -1/5\text{Cov}\left(N_{b_1,i}/m,N_{b_1,j}/m\right)$ and 
% \begin{align}
%     \eta&=\frac{m(m-1)(m-2)}{m^3n^2}+\frac{m(m-1)}{m^3n^2}\left(\frac{2}{m}-3\right)+\frac{2}{mn^3}+\frac{4-n}{n^3}\nonumber\\
%     &=\frac{7}{m^2n^2}-\frac{6}{mn^2}-\frac{2}{m^3n^2}+\frac{2}{mn^3}+\frac{4}{n^3}=-\frac{6}{mn^2}+\frac{4}{n^3}+\mcO(n^{-4}),\label{eq:eta}
% \end{align} that can be simplified to $\eta\approx 6\text{Cov}(N_{b_1,i}/m,N_{b_1,j}/m)+4\mbE[N_{b_1,i}/m]^3$. 

The unbiasedness of the proposed estimator, denoted as $ \widehat{\text{IF}}_2\left(D_i,D_j;\Fhat\right)$, is established in Theorem~\ref{thm:IFunbiased} under the assumption of the empirical distribution. 
\begin{theorem}\label{thm:IFunbiased}
Assuming $\Ybar^d(.)$ being a smooth function of input distribution and $Y^d_r(\Fhat^*_{b_1})$ are identically distributed and conditionally independent, $\widehat{\text{IF}}_2\left(D_i,D_j;\Fhat\right)$ defined in (\ref{eq:IF2}) is an unbiased estimator of $\nabla^2_{\Fhat}\theta$. 
\end{theorem} We can further expand the results of Theorem~\ref{thm:IFunbiased} and provide assurances regarding the convergence of the empirical CDF to the true distribution at an exponential rate \cite{massart1990empirical}. Particularly, as per initial assumptions $\theta$ is a smooth function of input distribution, therefore $\nabla^2\theta(.)$ is respectively a smooth function of the input distribution. Given $\nabla_{\Fhat}^3\theta(.)<\infty$, we can apply the delta method and Glivenko-Cantelli theorem \citep{loeve1977elementary} and achieve,\begin{align}
    \mbP\left\{\lim_{n\to\infty}\sup_{D_i,D_j}\bigg|\nabla^2_{\Fhat}\theta(D_i,D_j)-\nabla^2_{F}\theta(D_i,D_j)\bigg|=0\right\} =1.\nonumber
\end{align} 

Hence the bias estimate can be written as, \begin{align}
    \mbE_*[\theta(\Fhat^*)-\theta(\Fhat)]&=\mbE_*\left[\frac{1}{2}\sum_{i=1}^n\sum_{j=1}^n\widehat{\text{IF}}_2(i,j;\Fhat)\left(\frac{N_{b_1,i}}{m}-\frac{1}{n}\right)\left(\frac{N_{b_1,j}}{m}-\frac{1}{n}\right)\right]\nonumber\\
    &= \frac{1}{2}\mbE_*\Bigg[\sum_{i=1}^n\sum_{j=1}^n \frac{1}{R}\sum_{r=1}^{R} Y^d_r\left( \Fhat^*\right)S^{(2)}_{i,j}\left(\Fhat^*\right)\frac{S^{(2)}_{i,j}\left(\Fhat^*\right)}{\lambda}\nonumber\\
    &\ \ \ +\frac{S^{(2)}_{i,j}\left(\Fhat^*\right) Y^d_r(\Fhat)}{mn^2}-\eta S^{(2)}_{i,j}\left(\Fhat^*\right)\widehat{\text{IF}}_1\left(D_i;\Fhat\right)\Bigg]\nonumber\\
    &= \frac{\lambda}{2}\sum_{i=1}^n\sum_{j=1}^n\mbE_*\left[ \frac{1}{R}\sum_{r=1}^{R} Y^d_r\left( \Fhat^*\right)\left(\frac{N_{b_1,i}}{m}-\oon\right)^2\left(\frac{N_{b_1,j}}{m}-\oon\right)^2\right]\nonumber\\
    &= \frac{\lambda}{2}\sum_{i=1}^n\sum_{j=1}^n \text{Cov}_*\left(\frac{1}{R}\sum_{r=1}^{R} Y^d_r\left( \Fhat^*_{b_1}\right), \left(\frac{N_{b_1,i}}{m}-\oon\right)^2\left(\frac{N_{b_1,j}}{m}-\oon\right)^2\right)+\mcO(m^{-2}n^{-4}).\label{eq:HighVarBias}
\end{align} 

We can show that as $n$ goes to infinity, the variance of $\widehat{\text{IF}}_2$ becomes unbounded (Var$(\widehat{\text{IF}}_2)=\mcO(n^5)$). This means that in smaller datasets, we achieve a more stable estimator of bias than in larger datasets, which is not detrimental as the bias decreases with more data points. However, we can further reduce the variance by utilizing \emph{the control variate} technique. We use the $\widehat{\text{IF}}_1$ as the control variate statistic, since its expectation and variance are known. \emph{The control variate is our novelty in enhancing the HOIF bias estimator} from its original variation in \cite{Efron2014estaccuracy}.  It remains to compute the optimal coefficient of variance, that is equal to $c_2^*=-\text{Cov}(\widehat{\text{IF}_2}(\Fhat),\widehat{\text{IF}_1}(\Fhat))/\text{Var}(\widehat{\text{IF}_1}(\Fhat))$.
\begin{lemma}\label{thm:cov}
The optimal coefficient of variance $c_2^*=\frac{-\text{Cov}(\widehat{\text{IF}_2}(\Fhat),\widehat{\text{IF}_1}(\Fhat))}{\text{Var}(\widehat{\text{IF}_1}(\Fhat))}$ can be approximate with $-0.2\left(n-\frac{1}{m}+\frac{2}{mn}+1\right)+\mcO(m^{-2}n^{-2})$ due to 
\begin{align}
    \text{Cov}_*(\widehat{\text{IF}_2}(\Fhat),\widehat{\text{IF}_1}(\Fhat))&=\frac{\Ybar(\Fhat)^2}{5}\left(2+mn-n+mn^2+\frac{6}{mn}-\frac{6}{mn^2}-\frac{4}{n^2}+\frac{4}{n^3} \right).\nonumber\end{align}
    Here we drop the $D_i$ and $D_j$ from the influence function estimators to generalize the results for any point.
\end{lemma} 

% Hence the $c^*$ is\begin{align}
%     c^*&=\frac{-\text{Cov}(\widehat{\text{IF}_2},\widehat{\text{IF}_1})}{\text{Var}(\widehat{\text{IF}_1})}\approx -0.2\left(n-\frac{1}{m}+\frac{2}{mn}+1\right)+\mcO(m^{-2}n^{-2})
% \end{align}

% Proof of variance of $\widehat{\text{IF}_2}$:
% \begin{align}
%     \mbE[\widehat{\text{IF}}^2_2|\Fhat]&=\mbE[(\Ybar S_{i,j}+\frac{\lambda}{mn^2} Y-\lambda\eta\Ybar S_i)^2]\nonumber\\
%     &= \mbE[\Ybar^2S_{i,j}^2+\frac{\lambda^2}{m^2n^4} Y^2+\lambda^2\eta^2S_i^2\Ybar^2+2\Ybar S_{i,j}\frac{\lambda}{mn^2} Y-2\Ybar^2S_{i,j}\lambda\eta S_i-2\frac{\lambda^2\eta}{mn^2} Y\Ybar S_i],\nonumber
% \end{align}note that this expectation is with respect to the input distribution, not the simulation so, we can bring all $\theta^2$ out of every term, which results in, 
% \begin{align}
%     \mbE[\widehat{\text{IF}}^2_2|\Fhat]&=\theta^
% \end{align}
Following these results, the final bias estimator becomes for a given $b_1$ is \begin{align}
    \betahat(\Fhat)&\approx \left(-1.2\text{Cov}(N_{b_1,i}/m,N_{b_1,j}/m)\right)\nonumber\\
    &\ \ \ \times\sum_{i=1}^n\sum_{j=1}^n \text{Cov}_*\left(\frac{1}{R}\sum_{r=1}^{R} Y^d_r\left( \Fhat^*_{b_1}\right), \left(\frac{N_{b_1,i}}{m}-\oon\right)^2\left(\frac{N_{b_1,j}}{m}-\oon\right)^2\right).\label{eq:HOIFBias}
\end{align} 
\begin{remark}
    Viewing each bootstrapped sample as one simulation replication, the expected bias in (\ref{eq:HighVarBias}) is the same as second order linear regression coefficient of outputs with respect to the empirical cdf. For more intuition refer to \cite{lin2015single}.
\end{remark} 

We next prove in Theorem~\ref{thm:consistency} that the total bias estimator, $\betahat(\Fhat)+\What(\Fhat^*),$ converges to the true bias with probability 1, as the computation budget increases. Furthermore, Theorem~\ref{thm:consistency} shows that similar convergence results can be achieved with the variance-reduced FIB bias estimator, introduced in (\ref{eq:biasCVFIB}). 
\begin{theorem}\label{thm:consistency}
    The variance-reduced total bias estimator defined as $\What^{\text{cv}}(\Fhat^*)+\betahat(\Fhat)$ and total bias estimator, namely $\What(\Fhat^*)+\betahat(\Fhat)$, are unbiased estimators of $W(\Fhat^*)+\beta(\Fhat),$ i.e.,
    \begin{align}
        \lim_{N\to\infty}\mbE\left[|\What(\Fhat^*)+\betahat(\Fhat)-W(\Fhat^*)-\beta(\Fhat)|\right]=\lim_{N\to\infty}\mbE\left[|\What^{\text{cv}}(\Fhat^*)+\betahat(\Fhat)-W(\Fhat^*)-\beta(\Fhat)|\right]=0,\nonumber
    \end{align}where $N=B_1RB_2$ is the total computation budget. Furthermore, the proposed estimators converge to the true value as $N\to\infty,$ specifically,\begin{align}
        \mbE\left[Y(\Fhat^*)-\betahat(\Fhat)-\What(\Fhat^*)-\theta(F_0)\right]&\leq \mbE\left[Y(\Fhat^*)-\betahat(\Fhat)-\What^{\text{cv}}(\Fhat^*)-\theta(F_0)\right]\nonumber\\
        &=\mcO(\frac{1}{n^3})\nonumber
    \end{align}
\end{theorem}
The theorem above provides evidence that the proposed point estimator of the model output exhibits an error on the order of $n^{-3},$ indicating superior accuracy compared to the biased estimators. This improved precision in estimating the point estimator subsequently enhances the likelihood of achieving true coverage. 

\subsection{Debiased Confidence Intervals}\label{sec:DebCI}
Building upon the findings so far, we propose our debiased confidence intervals for the model output in Theorem~\ref{thm:propCI}.
\begin{theorem}\label{thm:propCI}
Let $\widehat{\text{Var}}\left(\Ybar^d(\Fhat^*)\right)$ follows \eqref{eq:YDebTotVar}.
    % Let, \begin{align}
    %     \widehat{\text{Var}}\left(\Ybar^d(\Fhat^*)\right)&=\frac{1}{B_1-1}\sum_{b_1=1}^{B_1}\left(\Ybar(\Fhat^*_{b_1})-\Ybar\right)^2+\frac{1}{RB_1(R-1)}\sum_{b_1=1}^{B_1}\sum_{r=1}^R\left(Y^d_r(\Fhat^*_{b_1})-\Ybar^d(\Fhat^*_{b_1})\right)^2 \nonumber \\ 
    %     &\ \ \ - \frac{1}{RB_1(R-1)}\sum_{b_1=1}^{B_1}\sum_{r=1}^R\left( Y_r(\Fhat^*_{b_1})-\Ybar(\Fhat^*_{b_1})\right)^2,\label{eq:totalDebVar}
    % \end{align} where $\Ybar^d(\Fhat^*_{b_1})=\frac{1}{R}\sum_{r=1}^R Y^d_r(\Fhat^*_{b_1}).$ 
    Then, assuming standard regularity conditions, we have \begin{align}
        \lim_{N\to\infty} \mbP\left\{ L_{\text{min}}(\alpha)\leq \theta(F_0) \leq L_{\text{max}}(\alpha)\right\} = 1-\alpha,\nonumber
    \end{align}where \begin{align}
        L_{\text{max}}(\alpha) = \frac{1}{B_1}\sum_{b_1=1}^{B_1}\Ybar^d(\Fhat^*_{b_1})+t_{RB_1-1,\alpha/2}\sqrt{\widehat{\text{Var}}\left(\Ybar^d(\Fhat^*_{b_1})\right)/(B_1R-1)},\nonumber
    \end{align}and \begin{align}
        L_{\text{min}}(\alpha) = \frac{1}{B_1}\sum_{b_1=1}^{B_1}\Ybar^d(\Fhat^*_{b_1})-t_{RB_1-1,\alpha/2}\sqrt{\widehat{\text{Var}}\left(\Ybar^d(\Fhat^*_{b_1})\right)/(B_1R-1)},\nonumber
    \end{align}where $t_{RB_1-1,\alpha/2}$ is the $\alpha/2$ quantile of student's t distribution with $RB_1-1$ degrees of freedom. 
\end{theorem}
The abovementioned theorem leverages the asymptotic distributional properties given by the CLT and the LLN to construct the confidence intervals. Even in scenarios with limited data, our CI offers the advantage of a reduced error in estimating the midpoint, thereby enhancing the probability of encompassing the true expected output. It is important to note that our CI's half-width is slightly broader compared to the biased CI. This is primarily attributed to the additional variability introduced when estimating the bias. However, it is worth mentioning that we have made efforts to minimize this variability through our variance-reduced variations of bias estimation.
Note that the results discussed here apply to both bias estimators, irrespective of whether the control variate technique is utilized or not.
In section~\ref{sec:num}, we thoroughly compare the proposed confidence intervals in Theorem~\ref{thm:propCI} and the existing state-of-the-art methods.

\subsection{Optimal Allocation}
\label{sec:alloc}
Fixing the simulation effort $N=B_1RB_2$, with \emph{the goal of having the simulation effort independent of the dataset size ($n$)}, one can find the optimum allocation of resources. \cite{LamQian2021} finds the best allocation for a nested simulation problem where sub-sampling has been incorporated into the outer simulation level. They prove that the optimum in the sense of minimizing the mean squared error of variance estimator is,\begin{align}
    \begin{cases} m^*=\Theta(N^{1/3}) &\text{if } 1\ll N\leq n^{3/2}\\
     \Theta(\sqrt{n})\leq m^*\leq \Theta(\max(1,N/n)) &\text{if } N>n^{3/2}
    \end{cases},\label{eq:LamOptm}
\end{align} which is translated,  in our case, to $R^*B^*_2=\Theta(m^*), B^*_1=\frac{N}{R^*B^*_2}$.

We further complete the analysis by finding the optimum allocation by minimizing the variance of the bias estimator introduced in Section~\ref{sec:bias}. The conditional variance of the simulation bias for arbitrary $b_2$ and $b'_2$ is\begin{align}
    \text{Var}_*\left(\frac{1}{B_2}\sum_{b_2=1}^{B_2}  Y_r\left(\Fhat_{b_1,b_2}^{**}\right)-  Y_r\left(\Fhat_{b_1,b_2}^{***}\right)\right)&=\frac{1}{B_2}\text{Var}_*\left( Y_r\left(\Fhat_{b_1,b_2}^{**}\right)-  Y_r\left(\Fhat_{b_1,b_2}^{***}\right)\right)\nonumber\\
    &\ \ \ -\frac{B_2-1}{B_2}\text{Cov}\left( Y_r\left(\Fhat_{b_1,b_2}^{**}\right)\right.\nonumber\\
    &\ \ \ \ \ \ \ \left.-  Y_r\left(\Fhat_{b_1,b_2}^{***}\right), Y_r\left(\Fhat_{b_1,b'_2}^{**}\right)-  Y_r\left(\Fhat_{b_1,b'_2}^{***}\right)\right),\nonumber
    %&=2\text{Cov}_*(\frac{1}{B_2}\sum_{b_2=1}^{B_2}  Y_r(\Fhat_{b_1,.,b_2}^{***}),\frac{1}{B_2}\sum_{b_2=1}^{B_2}  Y_r(\Fhat_{b_1,.,b_2}^{****})).\nonumber%\\
    %&=\frac{r^2m^*}{nB_2}\text{Var}(\theta)+\frac{r^3m^*}{nB_2}\text{Var}(\theta)-?
    %&=\frac{\text{Var}_*( Y_r(\Fhat_{b_1,.,b_2}^{***})- Y_r(\Fhat_{b_1,.,b_2}^{****}))}{B_2}\nonumber\\
   % &=\frac{n\sigma_{T}^2/(r^2m^*)}{B_2}=\Theta(1),\nonumber
\end{align}which can be rephrased as a function of $\text{Var}\left( Y\left(\Fhat^*_{b_1}\right)\right)$ using the sample variance of  bootstraps.
\begin{theorem}\label{thm:varbias}
Variance of the bias estimator in Section~\ref{sec:bias} is
\begin{align}
    \text{Var}\left(W_r(\Fhat^*_{b_1})\right)&= \frac{\sigma^2(2m^*-1)}{B_2(m^*)^2}\left(1+\frac{1}{B_2}\right)+\frac{(B_2-1)\sigma^2}{B_2m^*}\left(1+\frac{B_2-1}{B_2}\right)-2\frac{\sigma^2}{m^*}.\nonumber
\end{align} Ensuring that the variance of bias is in the same order as the variance of the sample average of bootstrap replications of the simulation results in,\begin{align}
    B_2^*=\Theta\left(\left(\frac{3(m^*)^2-m^*}{m^*+1}\right)^{1/3}\right),\nonumber
\end{align} subsequently $R^*=\Theta(\frac{m^*}{B_2^*})$.%\kimia{Then the error of the estimator becomes of order $n^{-4}$? }
\end{theorem}

%Note that as $B_2$ increases this variance decrease, as all terms correlate to the reciprocal of $B_2$. Hence, setting the largest feasible value for $B_2$ given the budget and other constraints ($R\geq 2$), we have \begin{align}
 %   B^*_2=\theta(\frac{m^*}{2}), \quad R^*=2.
%\end{align} 
% It follows that $B_2=\Theta(n/(r^2m^*))$ resulting in $R=\Theta(m^*/B_2)$. To maintain the lower bound for $R\geq 2$, we achieve\begin{align}
%    \frac{m^*m^*r^2}{n}\geq 2 \implies r\geq \sqrt{2n}/m^*.
%\end{align} Furthermore, we know $r<1$, hence, $\sqrt{2n}/m^*<1$ which results $m^*>\sqrt{2n}$.
\section{Notes on Machine Learning Output Analysis}
\label{sec:ML} 
A critical class of data-driven problems where the proposed method is directly applicable is predictive ML models. In building a ML model, the model's performance dramatically depends on the dataset used. For instance, if the training dataset is noisy and future data observations vary significantly from the input data, most ML models fail to achieve good prediction accuracy. Practitioners use different data-splitting measures to capture the conditional performance of ML models on the training data. In this section, we define a data sampling and splitting procedure derived from nested simulation settings in simulation \citep{SunApley2011nestedSim} that results in bias-corrected CI. 

We define the model output $Y(\cdot)$ as the squared error between the ML model prediction and the observed response for all data points being predicted. The error must be computed on points not used for model training to avoid overfitting bias \citep{ShashaaniVahdat2022}. 
%We define the process of generating one estimate of the ML model's performance as the simulation model \citep{ShashaaniVahdat2022}. 
Here nested simulation refers to iterated data sampling defined in Sections~\ref{sec:var} and \ref{sec:bias} for estimating the output bias and variance. Using each sampled distribution, $\Fhat^*_{b_1}$, one model is fit, and its error on another set is calculated as the desired output. The nested simulation needs identically distributed (i.d.) and consistent model accuracy estimates. To obtain i.d. replicates, we define $\Fhat^*_{b_1}$ as the distribution of the sub-samples, i.e., $m$-out-of-$n$ bootstrapping, with replacement for $b_1=1,\cdots,B_1$ \citep{Shao1996Boot}. As shown in the ML and statistics literature \citep{breiman2001random,Rabbi2020LinRegBoot}, using out-of-bag samples improves the model accuracy estimation and decreases the bias. We combine the two methodologies, where the model's performance is estimated on a $m$-out-of-$n$ bootstrap sample evaluated on the out-of-bag points. 

In the proposed algorithm, the model output, $Y$, is calculated via a weighted average of squared residuals. We denote the ML model and the dataset it is trained on with $h(\bm{D}),$ where $\bm{D}$ is the train data. Also, to show the predicted value of the trained model on a data point, $D_i$, we use $h(\bm{D}, D_i)$. Having in mind Figure~\ref{fig:frame} and \ref{fig:biasEst}, that demonstrate the required multi-level sampling for estimating bias and variance, we summarize our ML sampling algorithm in Algorithm~\ref{alg:main}. Each output shown on Figure~\ref{fig:biasEst} requires a pair of non-overlapping train and test sets. %We use the input distribution of $Y$ for testing, as the model error is calculated on them, and the holdout for training the model. 
\begin{algorithm}
   \caption{Sampling Algorithm For ML Error Estimation}
         \textbf{Given:} ML learning algorithm, dataset $\bm{D}$ of $n$ unique data points, simulation budget $N$.\\%, bootstrap size $m<n$ with $\kappa=m/n$.
         Compute the optimal allocated budget for $B_1^*, R^*, $ and $B_2^*$, and sub-sampling size $m^*$ using $N$, (\ref{eq:LamOptm}), and Theorem~\ref{thm:varbias}. Set $\kappa=m^*/n$.
      \\ \For{ sample $b_1=1,2,\cdots,B^*_1$}{
          Draw $\bm{D}_{b_1}$ from $\bm{D}$ with replacement and size $m^*$.\\
          Define $N_{b_1,i}=\#\{\bm{D}_{b_1}=D_i\}\ \forall i = 1,2,\cdots,n.$
       \\ \For{ simulation $r=1,2,\cdots,R^*$}{
          Draw a test sample $\bm{D}_{b_1,r}$ from $\bm{D}_{b_1}$. 
          Train model $h$ on $\bm{D}_{b_1,(r)}=\bm{D}_{b_1}\backslash\bm{D}_{b_1,r}$.\\
          Define $N_{b_1,r,i}=\#\{\bm{D}_{b_1,r}=D_i\},$ and $I_{b_1,r,i}=\mbI(D_{b_1,r,i}\in \bm{D}_{b_1,r} \And D_{b_1,r,i}\not\in \bm{D}_{b_1,(r)}) \forall i.$\\
          Calculate the model output with \begin{align}
             Y_r(\Fhat^{*}_{b_1})=\frac{1}{\sum_{i=1}^n I_{b_1,r,i}N_{b_1,r,i}}\sum_{i=1}^n I_{b_1,r,i}N_{b_1,r,i}\left(D_{b_1,r,i}-h(\bm{D}_{b_1,(r)}, D_{b_1,r,i})\right)^2.\label{eq:YEst}
         \end{align}
           \textbf{Bias Estimation:}
        \\\For{ bootstrap $b_2=1,\cdots,B^*_2$}{
                 Draw $\bm{D}_{b_1,r,b_2}$ from $\bm{D}_{b_1,r}$ of the size $\kappa\times|\bm{D}_{b_1,r}|$.
                 Train the model $h(\bm{D}_{b_1,r,(b_2)})$ whose outputs on $\bm{D}_{b_1,r,b_2}$ are $Y_{r}(\Fhat^{**}_{b_1,b_2})$ following (\ref{eq:YEst}).
                 Draw $\bm{D}_{b_1,r,b_2,1}$ from $\bm{D}_{b_1,r}$ of the size $\kappa^2\times|\bm{D}_{b_1,r}|$.\\
                 Train the model $h(\bm{D}_{b_1,r,(b_2),1})$ whose outputs on $\bm{D}_{b_1,r,b_2,1}$ are  $Y_{r}(\Fhat^{***}_{b_1,b_2})$ following (\ref{eq:YEst}).
                    }
                 Compute the debiased output, $Y^d_r(\Fhat^*_{b_1}),$ with (\ref{eq:biasFIB}).\\
                 Check for significance of bias estimate via the t-statistic defined in Theorem \ref{thm:convergance}.
     }
    }
     \textbf{Return:} The IU-induced bias with (\ref{eq:HOIFBias}), total variance with (\ref{eq:VarANOVA}), and the bias-corrected CI. \\
\label{alg:main}
\end{algorithm}

% \begin{enumerate}
%     \item For a specified ML model, $M$, take bootstrap sample with size $m\leq n$ from the whole data, $\bm{D}$.
% \end{enumerate} 
% Let $\bm{D}_{b_1,r}$ and $\bm{D}_{b_1,(r)}$ denote the sample taken with size $m$ from $\bm{D}_{b_1}$ and the left out observations in $\bm{D}_{b_1}\backslash \bm{D}_{b_1,r}$ for $1\leq m \leq n$, respectively. We define the simulation output as,\begin{align}
%      Y_r(\Fhat^{*}_{b_1})&=\mbE_{\bm{D}_{b_1,r}\sim \Fhat^{**}_{b_1,r}}\left[\left(\bm{D}_{b_1,r}-M_{\bm{D}_{b_1,(r)}}(\bm{D}_{b_1,r})\right)^2\right]\nonumber\\
%     &=\frac{1}{\sum_{i=1}^n I_{b_1,r,i}N_{b_1,r,i}}\sum_{i=1}^n I_{b_1,r,i}N_{b_1,r,i}\left(\bm{D}_{b_1,r,i}-M_{\bm{D}_{b_1,(r)}}(\bm{D}_{b_1,r,i})\right)^2, \label{eq:estimator1}
% \end{align} where \begin{align}
%     I_{b_1,r,i}=\mbI(\bm{D}_{b_1,r,i}\in \bm{D}_{b_1,r} \And \bm{D}_{b_1,r,i}\not\in \bm{D}_{b_1,(r)}),\text{ and } N_{b_1,r,i}=\#\{\bm{D}_{b,r}=D_i\}.\nonumber\end{align} 
    
   % In (\ref{eq:estimator1}), we calculate the weighted average of model error over the sub-sampled set. 
In Algorithm~\ref{alg:main}, we use the first random sample for testing the model, rather than building the model, which is essential in maintaining identically distributed error estimates for each data point \citep{ShashaaniVahdat2022}. Taking the first sample as the testing set results in having conditionally independent model performance estimates. Further, we maintain the sub-sampling ratio  $\kappa=m^*/n$ throughout to ensure optimized computational efficiency (see Section~\ref{sec:alloc}). 

\section{Numerical Experiments}
\label{sec:num}
This section demonstrates the applicability of the proposed method in machine learning. % and simulation settings. 
The ML case study estimates the model prediction mean squared error on the test set. We use three simulated datasets, where the true model error is known. In this experiment various methods in estimating the model error are compared with each other. The role of the input data size and optimal budget allocation introduced in section~\ref{sec:alloc} are studied. %The simulation application is shown on further analysis on the illustration example discussed in section~\ref{sec:ill}. 

%\subsection{Machine learning use case}\label{sec:numML}
We evaluate the performance of the proposed output analysis method by letting the under study functional be a machine learning regression model, where the purpose is to provide a correct confidence interval for the model mean squared error estimate. We conduct the experiment using three simulated data generating functions with two levels of noise added to the response to capture different functional structures and difficulty levels %, we experiment with a linear, polynomial, and a complex regression 
(see Table~\ref{tab:FuncDef}). In our analysis for ML type problem, the learning algorithm or model choice has no restriction, although we use linear regression in this section. Our main reason for using linear regression is to make the computation less costly. We also will show that although linear regression may not be a good choice for the underlying system logic, with reliable output analysis, it can still successfully predict good/bad alternatives.

\begin{table}
\TABLE{Description of the data generating functions. The noise added to each function, $\epsilon$, follows Normal distribution with mean zero and standard deviation 3 in the low noise and standard deviation 6 in the high noise cases. All independent variables, $z_1, z_2, $ and $z_3$ follow gamma distribution with shape parameters 2, 5, and 3 and rate parameters 1, 2, and 1, respectively. The simulated datasets have 100 observations. \label{tab:FuncDef}}{\begin{tabular}{@{}l|c|c|cc@{}}\toprule\centering

\textbf{Regression Type}&  Linear &  Polynomial & Complex \\ \midrule
\textbf{Formula} & $y=5z_1+z_2+2z_3+\epsilon$ & $y=z_1^2+z_1\times z_2+z_3^2+\epsilon$ & $y=5/\sqrt{z_1}+z_2+1/z_3+\epsilon$\\
\bottomrule
\end{tabular}}{}
\end{table}

We conduct a comparison between our suggested confidence interval when optimal budget allocation and variance reduction techniques (as discussed in Section~\ref{sec:bias}) are incorporated, and the confidence interval without these enhancements. Additionally, we compare our proposed approach with common sampling-based methods utilized in practice, namely Leave-One-Out Bootstrap (LOOBoot) and Repeated Cross-Validation. Further, we add two IU inflated confidence intervals in the literature to the comparison \citep{barton2018revisiting,LamQian2019}. Because there are no non-parametric bias estimators in the simulation (to the best of our knowledge), neither method considers bias in their CI. \cite{barton2018revisiting} estimates the variance similar to our nested simulation framework introduced in Section~\ref{sec:var}. Barton's method contrasts the impact of including the bias and its variance in the CI compared with our proposed method. The method proposed by \cite{LamQian2019} uses the first-order influence function, as described in (\ref{eq:IF1}), to estimate the IU variance. Interestingly, in most cases, Lam and Qian's method estimate the variability significantly larger than Barton's, which helps it to contain the actual value with more likelihood. 
\begin{figure}
\centering
\includegraphics[width=\textwidth]{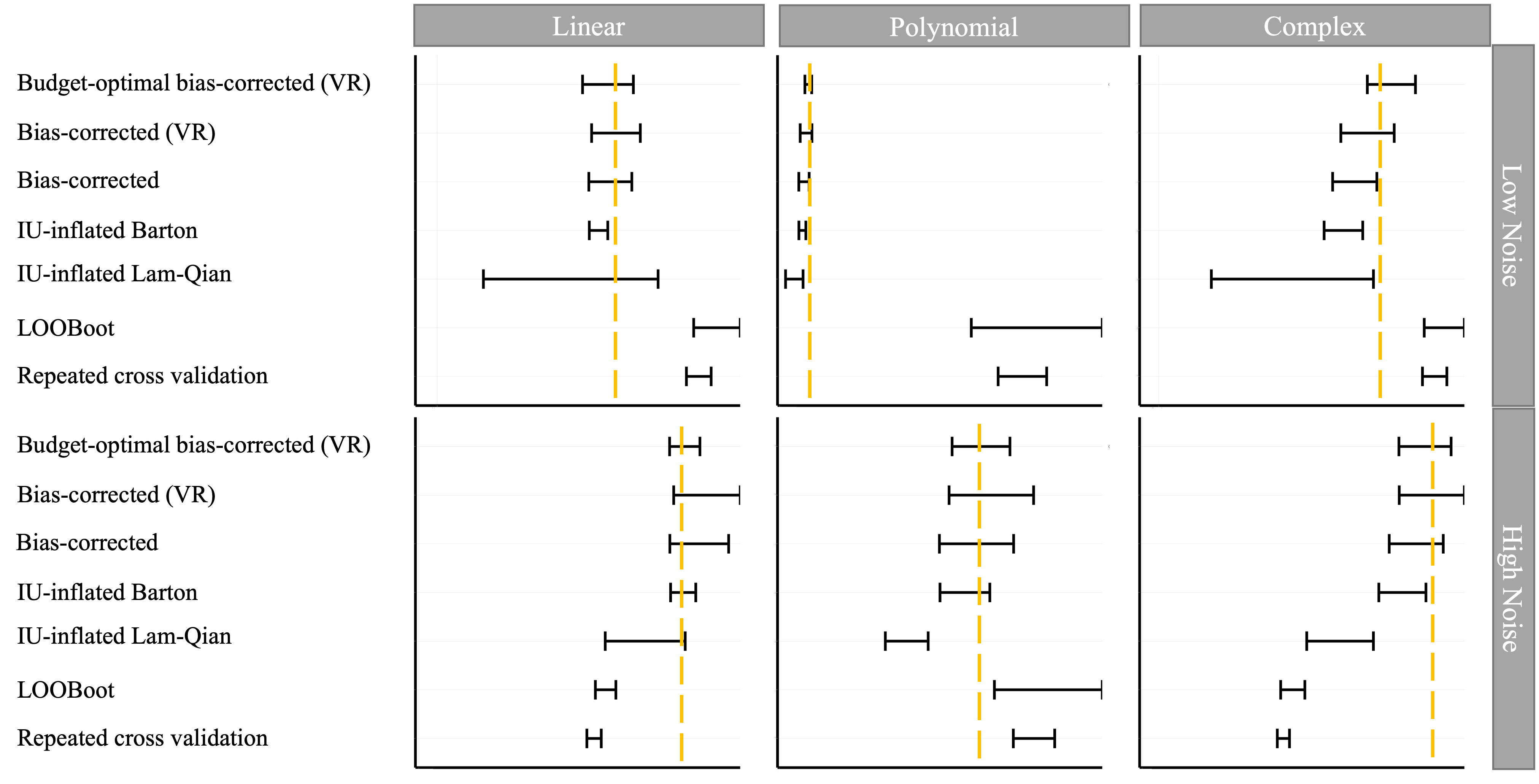}
\caption{Each panel shows the $95\%$ CI for the competing methods for different data generating functions and rows are for high in each column. The dashed lines show the expected value.%\sara{change CV to VR and font to Times New Roman: Done}
}
\label{fig:MLApp}
\end{figure}

% We compare our proposed confidence interval with and without optimal budget allocation with sampling-based methods commonly used in practice, leave one out bootstrap (LOOBoot), and repeated cross-validation (Repeated CV). Further, we add two IU inflated confidence intervals in the literature to the comparison \citep{barton2018revisiting,LamQian2019}. Because there are no non-parametric bias estimators in the simulation (to the best of our knowledge), neither method considers bias in their CI. \cite{barton2018revisiting} estimates the variance similar to our nested simulation framework introduced in Section~\ref{sec:var}. Barton's method contrasts the impact of including the bias and its variance in the CI compared with our proposed method. The method proposed by \cite{LamQian2019} uses the first-order influence function, as described in (\ref{eq:IF1}), to estimate the IU variance. Interestingly in most cases, Lam and Qian's method estimate the variability significantly larger than Barton's, which helps it to contain the actual value with more likelihood. 

\begin{figure}
\centering
\includegraphics[width=0.5\textwidth]{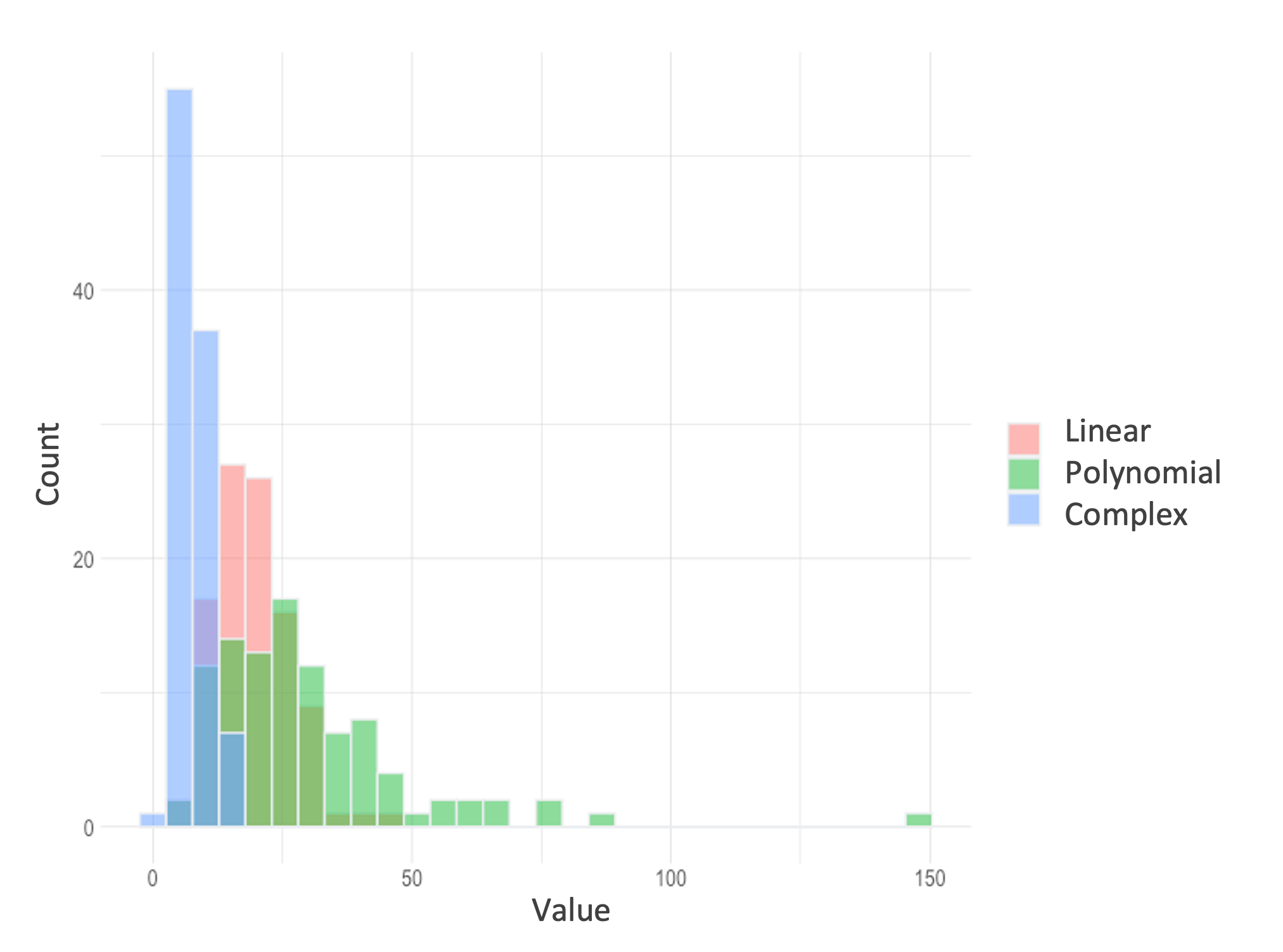}
\caption{Histogram of the response variable in simulated datasets. Distribution of the response variable encompasses the relationship and joint distributional properties of each dataset. The proposed algorithm builds CI for each case, independent from their parametric distributional attributes. The polynomial dataset in particular has a 
long tail, resulting in a more challenging prediction task. %\sara{explain this more clearly} 
}
\label{fig:Hist}
\end{figure}

To ensure a fair comparison, we maintain a fixed simulation budget of $N=1000$ across all cases. Table~\ref{tab:CovProb} presents a comparison of the coverage probability and confidence interval half widths for each method, considering multiple machine learning problems with two noise levels. The coverage probabilities are computed based on 100 replications. 
The confidence intervals for a single replication are illustrated in Figure~\ref{fig:MLApp}. To gain deeper insights into the performance variation among the simulated datasets, we can examine the distribution of the responses, as shown in Figure~\ref{fig:Hist}. The presence of a long tail in the distribution has made the polynomial data a more demanding prediction task. Consequently, the distinction between the debiased CIs and the biased competitors becomes more pronounced.
% We keep the simulation budget fixed to 1000 for all cases to provide a fair ground for comparison. Table~\ref{tab:CovProb} compares the coverage probability for each method over the mentioned ML problems with two noise levels. The coverage probabilities are computed over 100 replications. We can observe that the bias-corrected CI significantly outperform the competitors. Lam and Qian, and Barton's CI perform relatively better than repeated CV and LOOBoot in complex and linear functions. Moreover, we can improve the coverage probability with optimized budget allocation.
% Figure~\ref{fig:MLApp} demonstrates the CI for one replication.

% \begin{figure}
%     \FIGURE{\includegraphics[width=\textwidth]{./figs/MLexample.png}}{Each panel shows the $95\%$ CI for the competing methods for different data generating functions and rows are for high in each column. The dashed lines show the expected value.\label{fig:MLApp}}{}
% \end{figure}
\begin{table}\small
\centering
    \caption{The probability of true coverage and average CI half-width are calculated over 100 replications. %The first row is the proposed method with optimal budget allocation and ``(VR)" denotes the variance-reduced estimator as discussed in Theorem~\ref{thm:FIBCVunbiased}. The second row is the same proposed method without optimal budget allocation and the third row is the bias corrected estimator without variance reduction and budget allocation. 
    The best coverage and half-width in each experiment are shown in bold. %\sara{check ``?'', write cov in XX format}
    }
    \label{tab:CovProb}
    \begin{tabular}{@{}l|c|ccc|c|cc|cc|cc@{}}\toprule\centering
    & \multicolumn{4}{c|}{ } & Regression Type & \multicolumn{2}{c|}{Linear}  & \multicolumn{2}{c|}{Polynomial} & \multicolumn{2}{c}{Complex}   \\ 
    \hline
    & \multicolumn{4}{c|}{Budget} & & \multicolumn{6}{c}{Level of noise} \\ \cline{2-5}\cline{7-12}
    Output Analysis Method & Total & $B_1$ & $R$ & $B_2$& Performance & low & high & low & high & low & high \\ \midrule
    \multirow{2}{*}{Opt. Bias-corrected (VR)} & \multirow{2}{*}{1000} & \multirow{2}{*}{50} & \multirow{2}{*}{5} & \multirow{2}{*}{4} & Coverage \% & \textbf{100} & \textbf{100}  & \textbf{40}  & \textbf{27}  & \textbf{93}  & \textbf{95}  \\
    &  &&&& Half-Width &0.6 & 2.8 & \textbf{1.0} & 10.7 & 0.8 & 6.0 \\\hline
    \multirow{2}{*}{Bias-corrected (VR)} & \multirow{2}{*}{1000} & \multirow{2}{*}{10} & \multirow{2}{*}{10} & \multirow{2}{*}{10}& Coverage \% & 93 & 93 & 33 & \textbf{27}  & 90& 93 \\
    & &&&& Half-Width &0.6 & 5.3 & 1.5 & 13.7 & 0.7 & 6.2 \\\hline
    \multirow{2}{*}{Bias-corrected} & \multirow{2}{*}{1000} & \multirow{2}{*}{10} & \multirow{2}{*}{10} & \multirow{2}{*}{10}& Coverage \% & 83 & 83  & 33 & 10 & 30 & 90 \\
    & &&&& Half-Width&0.7 & 6.0 & 1.7 & 15.6 & 0.8 &7.5 \\\hline\hline
    \multirow{2}{*}{IU-inflated Barton} & \multirow{2}{*}{1000} & \multirow{2}{*}{100} & \multirow{2}{*}{10} & \multirow{2}{*}{-} & Coverage \% & 70 & 70 & \textbf{36} & 10 & 40 & 20 \\
    &&&&& Half-Width&\textbf{0.3} & 2.3 & \textbf{1.0} & 9.2 & 0.6 & 5.4\\\hline
    \multirow{2}{*}{IU-inflated Lam-Qian} & \multirow{2}{*}{1000} & \multirow{2}{*}{100} & \multirow{2}{*}{10} & \multirow{2}{*}{-} & Coverage \% & \textbf{100} & 10 & 0 & 0 & 0  & 0  \\
    &&&&&Half-Width&2.4 & 7.2 & 2.6 & 7.9 & 2.5 & 7.7 \\\hline
    \multirow{2}{*}{LOOBoot} & \multirow{2}{*}{1000} & \multirow{2}{*}{100} & \multirow{2}{*}{10} & \multirow{2}{*}{-} & Coverage \% & 15 & 0 & 0 & 0 & 5 & 2 \\
     &&&&&Hlaf-Width& 0.6 & 1.8 & 19.4 & 19.9 & 0.6 & 2.7 \\\hline
    \multirow{2}{*}{Repeated cross-validation} & \multirow{2}{*}{1000} & \multirow{2}{*}{100} & \multirow{2}{*}{10} & \multirow{2}{*}{-} & Coverage \% & 10  & 5  & 0  & 0  & 2 & 3 \\
     &&&&&Half-Width& \textbf{0.3} & \textbf{1.3} & 7.2 & \textbf{7.7}& \textbf{0.4}&\textbf{1.4}\\
    \bottomrule
    \end{tabular}
\end{table}

The results indicate that the bias-corrected and variance-reduced (shown with ``(VR)" in the table) confidence interval outperforms the other methods significantly. In complex and linear functions, Lam and Qian's CI and Barton's CI demonstrate relatively better performance compared to repeated cross-validation and LOOBoot. Furthermore, by employing optimized budget allocation, we can enhance the coverage probability, while reducing the CI's half width. It should be noted that the half width of our proposed confidence intervals is comparable to that of Barton's. This similarity stems from the fact that our variance estimators are computed in a similar manner, with a slight increase in variance due to the inclusion of multi-layer bias estimation in our confidence interval. Interestingly, repeated cross-validation yields the smallest half widths among all methods, leading to a near-zero probability of coverage. 
% \begin{table}
%     \TABLE{With the total simulation budget of 1000, the probability of correct coverage is calculated over 100 replications. The first row is the proposed method and the second row is the proposed method with optimal budget allocation. The best coverage in each experiment is bolded.\label{tab:CovProb}}{\begin{tabular}{@{}l|cc|cc|ccc@{}}\toprule\centering
%     \textbf{Regression Type}& \multicolumn{2}{c}{Linear}  & \multicolumn{2}{c}{Polynomial} & \multicolumn{2}{c}{Complex}  \\ \cline{2-7}
%     Noise Variance & low  & high  & low  & high  & low  & high  \\ \midrule
%     Budget-optimal bias-corrected  CI coverage& \textbf{0.83} & \textbf{0.85} & 0.31 & \textbf{0.15} & 0.79 & \textbf{0.92} \\
%     Bias-corrected CI coverage& 0.76 & 0.76 & \textbf{0.35} & 0.11 & 0.76 & 0.89 \\\hline
%     IU-inflated Barton  CI coverage& 0.33 & 0.33 & 0.00 & 0.06 & 0.36 & 0.10\\
%     IU-inflated Lam-Qian  CI coverage& 0.75 & 0.71 & 0.00 & 0.00 & \textbf{0.81} & 0.82  \\
%     LOOBoot  CI coverage&0.15 & 0.00 & 0.02 & 0.03 & 0.00 & 0.00 \\
%     Repeated CV  CI coverage& 0.10 & 0.05 & 0.02 & 0.0 & 0.00 & 0.00\\
%     \bottomrule
%     \end{tabular}}{}
% \end{table}

\section{Concluding Remarks}
\label{sec:conclusion}
This paper showed the significance of incorporating bias and variance estimation in the model output analysis. We emphasize this matter via two common practical problems of machine learning and stochastic simulation. We bridge between the two fields providing a new playground for future research in the interconnection of ML and simulation. 

We focused on non-parametric estimation methods to keep our results generalizable to data-driven problems. Furthermore, we addressed the computation inefficiency of non-parametric methods with an optimal budget allocation, which facilitates us to keep the computing budget the same while estimating the bias and variance of the model output. However, we do not include the model building cost in our computation budget, which can potentially be a drawback if a more complex model is fit. This problem can be addressed by restricting the non-parametric assumption to replace the bias estimator with a less number of resampling \citep{lin2015single}, which we leave for future research.

For both simulation and ML, our proposed method can be especially beneficial for big data problems, when due to computational expenses, the user can take smaller subsets of data for scalability. Leveraging the proposed bias-corrected CI can compensate for the loss of data. %\sara{might need to move to conclusion}

Viewing ML as a simulation clarifies the propagation of bias of data into output. Without prediction bias, the estimates of future outcomes of a decision can mislead the decision-maker into choosing a worse and riskier option. One of the future research paths of interest would be incorporating the proposed estimator into data-driven optimization problems.
% Acknowledgments here
\ACKNOWLEDGMENT{ The preliminary results of this paper was submitted to WSC 2021 \citep{Vahdat2021}. The authors are also thankful to the AAUW Research Publication Grant in Engineering, Medicine and Science, American Educational Research Association that partially funded this project.
}% Leave this (end of acknowledgment)

% Appendix here
% Options are (1) APPENDIX (with or without general title) or 
%             (2) APPENDICES (if it has more than one unrelated sections)
% Outcomment the appropriate case if necessary
%

%
%   or 
%
\begin{APPENDICES}
\section{Theorem Proofs}\label{sec:proofs}
\subsection{Proof of Theorem \ref{thm:convergance}}
\proof{proof:}
Recall that $\What_r(\Fhat^*_{b_1})=\bar{\Delta}^{*}_{r}\left(\Fhat^{**}_{b_1,.}\right)+\hat{\gamma}_r\left(\Fhat^{**}_{b_1,.}\right)$. Assume $\text{Var}\left(W_r(\Fhat^*_{b_1})\right)<\infty.$ By the central limit theorem (More in-depth discussion on asymptotic behavior and validity of CLT in optimization space can be found in  \cite{hunter2022central}.) and definition of $\What_r(\Fhat^*_{b_1})$, as $B_2$ grows larger, $\What_r(\Fhat^*_{b_1})$ converges in distribution to Normal distribution with mean $W_r(\Fhat^*_{b_1}),$ and variance $\text{Var}\left(W_r(\Fhat^*_{b_1})\right).$ Moreover, the summation of 
$l$ squared standard independent Normal variables has Chi-squared distribution with $l$ degrees of freedom. Hence \begin{align}
    \sum_{b_2=1}^{B_2}\left(\Delta^{*}_r(\Fhat^{**}_{b_1,b_2})+\hat{\gamma}_r(\Fhat^{**}_{b_1,b_2})-\What_{r}(\Fhat^*_{b_1})\right)^2\sim \chi^2 (B_2-1),\nonumber
\end{align} and $T$ follows student's t distribution with $B_2-1$ degrees of freedom.

To show the validity of the proposed confidence intervals, note that for some $\alpha\in(0,1)$ and as $B_2\to\infty$ \begin{align}
    &\mbP\Bigg\{ -t_{B_2-1,\alpha/2}\leq\frac{\bar{\Delta}_{r}^{*}(\Fhat^{**}_{b_1,.})+\hat{\gamma}_r(\Fhat^{**}_{b_1,.})-W_r(\Fhat^*_{b_1})}{\sqrt{\frac{1}{B_2-1}\sum_{b_2=1}^{B_2}\left(\Delta^{*}_{r}(\Fhat^{**}_{b_1,b_2})+\hat{\gamma}_r(\Fhat^{**}_{b_1,b_2})-\What_{r}(\Fhat^*_{b_1})\right)^2}}\leq t_{B_2-1,\alpha/2}\Bigg\}=1-\alpha, \nonumber\\
    \implies &\mbP\Bigg\{-t_{B_2-1,\alpha/2}\sqrt{\frac{1}{B_2-1}\sum_{b_2=1}^{B_2}\left(\Delta^{*}_{r}(\Fhat^{**}_{b_1,b_2})+\hat{\gamma}_r(\Fhat^{**}_{b_1,b_2})-\What_{r}(\Fhat^*_{b_1})\right)^2}\leq\nonumber\\ 
    &\ \ \ \ \ \bar{\Delta}_{r}^{*}(\Fhat^{**}_{b_1,.})+\hat{\gamma}_r(\Fhat^{**}_{b_1,.})-W_r(\Fhat^*_{b_1})\leq \nonumber\\
    &\ \ \ \ \ t_{B_2-1,\alpha/2}\sqrt{\frac{1}{B_2-1}\sum_{b_2=1}^{B_2}\left(\Delta^{*}_{r}(\Fhat^{**}_{b_1,b_2})+\hat{\gamma}_r(\Fhat^{**}_{b_1,b_2})-\What_{r}(\Fhat^*_{b_1})\right)^2} \Bigg\}=1-\alpha\nonumber\\
    \implies & \mbP\left\{T_{\text{min}}(\alpha)\leq W_r(\Fhat^*_{b_1})\leq T_{\text{max}}(\alpha)\right\}=1-\alpha.\nonumber
\end{align}
\endproof
\subsection{Proof of Theorem \ref{thm:IFunbiased}.}

%\begin{proof}
\proof{\textit{Proof.}}
We first begin by deriving the (\ref{eq:lambda}) and (\ref{eq:eta}), then show that given these definitions, the $\widehat{\text{IF}}_2\left(D_i,D_j;\Fhat\right)$ is conditionally unbiased. \begin{align}
    \frac{2}{\lambda}&=\frac{m(m-1)(m-2)(m-3)}{m^4n^2}+\frac{m(m-1)(m-2)}{m^3n^3}\left(\frac{5n}{m}-4n\right)\nonumber\\
    &\ \ \ +\frac{m(m-1)}{n^2m^2}\left(\frac{4}{m^2}+\frac{8}{mn}-\frac{8}{mn^2}+6\right)-\frac{4}{mn^3}-\frac{3}{n^2}-\frac{2}{m^3n}+\frac{5}{mn^2},\nonumber\\
    &=\frac{10}{mn^2}-\frac{8}{m^2n^2}+\frac{4}{mn^3}-\frac{8}{mn^4}-\frac{8}{m^2n^3}+\frac{8}{m^2n^4}-\frac{2}{m^3n}=\frac{10}{mn^2}+\mcO\left(n^{-4}\right).\nonumber
\end{align} Hence $\lambda\approx -1/5\text{Cov}\left(N_{b_1,i}/m,N_{b_1,j}/m\right)$ and 
\begin{align}
    \eta&=\frac{m(m-1)(m-2)}{m^3n^2}+\frac{m(m-1)}{m^3n^2}\left(\frac{2}{m}-3\right)+\frac{2}{mn^3}+\frac{4-n}{n^3}\nonumber\\
    &=\frac{7}{m^2n^2}-\frac{6}{mn^2}-\frac{2}{m^3n^2}+\frac{2}{mn^3}+\frac{4}{n^3}=-\frac{6}{mn^2}+\frac{4}{n^3}+\mcO(n^{-4}),\nonumber
\end{align} that can be simplified to $\eta\approx 6\text{Cov}(N_{b_1,i}/m,N_{b_1,j}/m)+4\mbE[N_{b_1,i}/m]^3$. 

Next, we need to show that the conditional expectation of the second order IF estimator given the empirical distribution is unbiased, i.e.,\begin{align}
    \mbE_Y\left[\widehat{\text{IF}}_2\left(D_i,D_j;\Fhat\right)|\Fhat\right]&= \mbE_Y\left[\frac{1}{B_1}\sum_{b_1=1}^{B_1}\frac{1}{R}\sum_{r=1}^{R} Y^d_r\left( \Fhat^*_{b_1}\right)S^{(2)}_{i,j}\left(\Fhat^*_{b_1}\right)+\frac{\lambda Y^d_r(\Fhat)}{mn^2}-\lambda\eta\widehat{\text{IF}}_1\left(D_i;\Fhat\right)\right]\nonumber\\
    &=\frac{1}{B_1}\sum_{b_1=1}^{B_1}\frac{1}{R}\sum_{r=1}^{R}\mbE_Y\left[ Y^d_r\left( \Fhat^*_{b_1}\right)S^{(2)}_{i,j}\left(\Fhat^*_{b_1}\right)+\frac{\lambda Y^d_r(\Fhat)}{mn^2}\right]-\mbE_Y\left[\lambda\eta\widehat{\text{IF}}_1\left(D_i;\Fhat\right)\right]\nonumber\end{align}
By substituting the $S^{(2)}_{i,j}\left(\Fhat^*_{b_1}\right)$ according to its definition, the above equation becomes equivalent to
    \begin{align}
    &\mbE\Bigg[Y^d_r(\Fhat)\lambda\left(\frac{N_{b_1,i}}{m}-\oon\right)\left(\frac{N_{b_1,j}}{m}-\oon\right)+\sum_{i'=1}^n\nabla_{\Fhat}\theta(z_{i'})\left(\frac{N_{i'}}{m}-\oon\right)\lambda\left(\frac{N_{b_1,i}}{m}-\oon\right)\left(\frac{N_{b_1,j}}{m}-\oon\right)\nonumber\\
    &+\frac{1}{2}\sum_{i'=1}^n\sum_{j'=1}^n\nabla^2_{\Fhat}\theta(z_{i'},z_{j'})\left(\frac{N_{i'}}{m}-\oon\right)\left(\frac{N_{j'}}{m}-\oon\right)\lambda\left(\frac{N_{b_1,i}}{m}-\oon\right)\left(\frac{N_{b_1,j}}{m}-\oon\right)\Bigg]+\frac{\lambda Y^d_r(\Fhat)}{mn^2}-\lambda\eta\nabla_{\Fhat}\theta\nonumber\\
    &=\lambda Y^d_r(\Fhat)\text{Cov}\left(\frac{N_{i}}{m},\frac{N_{j}}{m}\right)+\nabla_{\Fhat}\theta\lambda\mbE\left[\sum_{i'=1}^n\left(\frac{N_{i'}}{m}-\oon\right)\left(\frac{N_{b_1,i}}{m}-\oon\right)\left(\frac{N_{b_1,j}}{m}-\oon\right)\right]\nonumber\\
    &\ \ \ +\frac{1}{2}\nabla^2_{\Fhat}\theta\lambda\mbE\left[\sum_{i'=1}^n\sum_{j'=1}^n\left(\frac{N_{i'}}{m}-\oon\right)\left(\frac{N_{j'}}{m}-\oon\right)\left(\frac{N_{b_1,i}}{m}-\oon\right)\left(\frac{N_{b_1,j}}{m}-\oon\right)\right]+\frac{\lambda Y^d_r(\Fhat)}{mn^2}-\lambda\eta\nabla_{\Fhat}\theta.\nonumber\end{align}
Exploiting the distributional properties of $N_i,$ and knowing its higher order moments, allows us to further simplify the above expression to
    \begin{align}
    &\lambda Y^d_r(\Fhat)(\frac{-1}{mn^2})+\lambda\nabla_{\Fhat}\theta\left(\frac{m(m-1)(m-2)}{m^3n^2}+\frac{m(m-1)}{m^3n^2}\left(\frac{2}{m}-3\right)+\frac{2}{mn^3}+\frac{4-n}{n^3}\right)\nonumber\\
    &\ \ \ +\frac{1}{2}\nabla^2_{\Fhat}\theta\lambda\Bigg( \frac{m(m-1)(m-2)(m-3)}{m^4n^2}+\frac{m(m-1)(m-2)}{m^3n^3}\left(\frac{5n}{m}-4n\right)\nonumber\\
    &\ \ \ +\frac{m(m-1)}{n^2m^2}\left(\frac{4}{m^2}+\frac{8}{mn}-\frac{8}{mn^2}+6\right)-\frac{4}{mn^3}-\frac{3}{n^2}-\frac{2}{m^3n}+\frac{5}{mn^2}\Bigg)+\frac{\lambda Y^d_r(\Fhat)}{mn^2}-\lambda\eta\nabla_{\Fhat}\theta.\nonumber
\end{align} By replacing the $\lambda$ with (\ref{eq:lambda}) and $\eta$ with, we get, \begin{align}
    \mbE\left[\widehat{\text{IF}}_2\left(D_i,D_j;\Fhat\right)|\Fhat\right]=\frac{1}{2}\lambda\nabla^2_{\Fhat}\theta(D_i,D_j)\frac{2}{\lambda}=\nabla^2_{\Fhat}\theta(D_i,D_j).\nonumber%\label{eq:IF2unbiased}
\end{align} 
As per initial assumptions $\theta(.)$ is a smooth function of input distribution, therefore $\nabla\theta(.)$ is respectively a smooth function of the input distribution. Given $\nabla_{\Fhat}^2\theta(.)<\infty$, we can apply the delta method and Glivenko-Cantelli theorem \citep{loeve1977elementary} and achieve,\begin{align}
    \mbP\left\{\lim_{n\to\infty}\sup_{D_i,D_j}\bigg|\nabla^2_{\Fhat}\theta(D_i,D_j)-\nabla^2_{F}\theta(D_i,D_j)\bigg|=0\right\} =1.\nonumber
\end{align} 
\endproof
%\end{proof}

\subsection{Proof of Lemma \ref{thm:cov}.}
%\begin{proof}
\proof{\textit{Proof.}}
We take advantage of Covariance additivity property to calculate the covariance between $\widehat{\text{IF}}_2(\Fhat)$ and $\widehat{\text{IF}}_1(\Fhat)$. \begin{align}
    \text{Cov}_*\left(\widehat{\text{IF}}_2(\Fhat),\widehat{\text{IF}}_1(\Fhat)\right)&= \text{Cov}_*\left(\Ybar(\Fhat) S_{i,j}+\frac{\lambda}{mn^2}-\lambda\eta\Ybar(\Fhat) S_i, \Ybar(\Fhat) S_i\right)\nonumber\\
    & = \text{Cov}_*\left(\Ybar(\Fhat) S_{i,j}, \Ybar(\Fhat) S_i\right)+\frac{\lambda}{mn^2}\text{Cov}_*\left( Y(\Fhat),\Ybar(\Fhat) S_i\right)\nonumber\\
    &\ \ \ -\lambda\eta\text{Cov}_*\left(\Ybar(\Fhat) S_i,\Ybar(\Fhat) S_i\right)\nonumber\\
    &=\Ybar(\Fhat)^2(mn\lambda)\text{Cov}\left(\left(\frac{N_i}{m}-\oon\right)\left(\frac{N_j}{m}-\oon\right),\left(\frac{N_i}{m}-\oon\right)\right)-\lambda\eta\text{Var}\left(\Ybar(\Fhat) S_i\right)\nonumber\\
    &= \Ybar(\Fhat)^2(mn\lambda)\Bigg(\mbE\left[(\frac{N_i}{m}-\oon)^2(\frac{N_j}{m}-\oon)\right]\nonumber\\ 
    &\ \ \ - \mbE\left[(\frac{N_i}{m}-\oon)(\frac{N_j}{m}-\oon)\right]\mbE\left[(\frac{N_i}{m}-\oon)\right] \Bigg)- \Ybar(\Fhat)^2\lambda\eta\text{Var}\left(\frac{N_i}{m}-\oon\right).\nonumber
    \end{align} Now, given that the distribution of $N_i$ is known, we can simplify the above to,
    \begin{align}
    \text{Cov}_*\left(\widehat{\text{IF}}_2(\Fhat),\widehat{\text{IF}}_1(\Fhat)\right)&= \frac{\Ybar(\Fhat)^2mn(mn^2)}{5}\left(\frac{1}{mn}\left(\frac{2}{mn^2}+\oon-\frac{1}{mn}+1\right)\right)\nonumber\\
    &\ \ \ -\frac{\Ybar(\Fhat)^2mn^2}{5}\left(\frac{-6}{mn^2}+\frac{4}{n^3}\right)\left(\frac{1}{mn}-\frac{1}{mn^2}\right)\nonumber\\
    &= \frac{\Ybar(\Fhat)^2}{5}\left(2+mn-n+mn^2+\frac{6}{mn}-\frac{6}{mn^2}-\frac{4}{n^2}+\frac{4}{n^3} \right).\nonumber
\end{align} 
\endproof
%\end{proof}

\subsection{Proof of Theorem \ref{thm:consistency}.}
\proof{proof:}
The unbiased property follows directly from Lemmas~\ref{thm:FIBunbiased} and ~\ref{thm:FIBCVunbiased} and Theorem~\ref{thm:IFunbiased}. 
The debiased estimator can be expanded as the difference of the simulation output and the bias terms, i.e.,
\begin{align}
    \mbE[Y(\Fhat^*)-\betahat(\Fhat)-\What^{\text{cv}}(\Fhat^*)]&= \mbE[Y(\Fhat^*)-\beta(\Fhat)-W(\Fhat^*)]+\mbE[\beta(\Fhat)-\betahat(\Fhat)]+\mbE[W(\Fhat^*)-\What^{\text{cv}}(\Fhat^*)]\nonumber\\
    &=\theta(F_0)+\mcO(\frac{1}{n^3})+\mcO(\frac{1}{\sqrt{B_1R}}) +\mcO(\frac{1}{B_2})+\mcO(\frac{1}{R}).\nonumber
\end{align} The last equality is a direct derivation of (\ref{eq:Y-decompose2}) and unbiased properties of two bias estimators. 

\endproof

\subsection{Proof of Theorem \ref{thm:propCI}.}
\proof{proof:}
Assume $\text{Var}\left(Y^d_r(\Fhat^*_{b_1})\right)<\infty.$ 
% The employing the law of total variance and findings of section~\ref{sec:var}, we can write\begin{align}
%     \text{Var}\left(\Ybar^d(\Fhat^*) \right) &= \mbE_*\left[\text{Var}\left(\Ybar^d(\Fhat^*)|\Fhat^*\right)\right]+\text{Var}_*\left(\mbE\left[\Ybar^d(\Fhat^*)|\Fhat^*\right]\right).\nonumber
% \end{align} Applying (\ref{eq:VarANOVA}) and estimating the first term above with sum of squared errors, we can get an unbiased estimator for the total variance as follows,
% \begin{align}
%     \widehat{\text{Var}}\left(\Ybar^d(\Fhat^*)\right)&=\frac{1}{B_1-1}\sum_{b_1=1}^{B_1}\left(\Ybar(\Fhat^*_{b_1})-\Ybar\right)^2+\frac{1}{RB_1(R-1)}\sum_{b_1=1}^{B_1}\sum_{r=1}^R\left(Y^d_r(\Fhat^*_{b_1})-\Ybar^d(\Fhat^*_{b_1})\right)^2 \nonumber \\ 
%         &\ \ \ - \frac{1}{RB_1(R-1)}\sum_{b_1=1}^{B_1}\sum_{r=1}^R\left( Y_r(\Fhat^*_{b_1})-\Ybar(\Fhat^*_{b_1})\right)^2.\nonumber
% \end{align}
Similar to the proof of Theorem~\ref{thm:convergance}, by the law of large numbers and the central limit theorem, as $R\times B_1$ grows larger, $\Ybar^d(\Fhat^*)$ converges to Normal distribution and subsequently, \begin{align}
    \frac{\Ybar^d(\Fhat^*)-\theta(F_0)}{\sqrt{\widehat{\text{Var}}\left(\Ybar^d(\Fhat^*)\right)/(RB_1-1)}}\nonumber
\end{align} converges to student's t distribution with $RB_1-1$ degrees of freedom.

To show the validity of the proposed confidence intervals, note that for some $\alpha\in(0,1)$ and as $RB_1\to\infty$ \begin{align}
    &\mbP\Bigg\{ -t_{RB_1-1,\alpha/2}\leq\frac{\Ybar^d(\Fhat^*)-\theta(F_0)}{\sqrt{\widehat{\text{Var}}\left(\Ybar^d(\Fhat^*)\right)/(RB_1-1)}}\leq t_{RB_1-1,\alpha/2}\Bigg\}=1-\alpha, \nonumber\\
    \implies &\mbP\Bigg\{-t_{RB_1-1,\alpha/2}\sqrt{\widehat{\text{Var}}\left(\Ybar^d(\Fhat^*)\right)/(RB_1-1)}\leq \Ybar^d(\Fhat^*)-\theta(F_0)\leq t_{RB_1-1,\alpha/2}\sqrt{\widehat{\text{Var}}\left(\Ybar^d(\Fhat^*)\right)/(RB_1-1)} \Bigg\}=1-\alpha\nonumber\\
    \implies & \mbP\left\{L_{\text{min}}(\alpha)\leq \theta(F_0) \leq L_{\text{max}}(\alpha)\right\}=1-\alpha.\nonumber
\end{align}
\endproof
\subsection{Proof of Theorem \ref{thm:varbias}.}
%\begin{proof}
\proof{\textit{Proof.}} Using the law of total variance we have,
\begin{align}
    \text{Var}\left(\What_r(\Fhat^*_{b_1})\right)&=\text{Var}\left(\frac{1}{B_2}\sum_{b_2=1}^{B_2} Y_r\left(\Fhat^{**}_{b_1,b_2}\right)-\frac{1}{B_2}\sum_{b_2=1}^{B_2} Y_r\left(\Fhat^{***}_{b_1,b_2}\right)\right)\nonumber\\
    &=\text{Var}\left(\frac{1}{B_2}\sum_{b_2=1}^{B_2} Y_r\left(\Fhat^{**}_{b_1,b_2}\right)\right) + \text{Var}\left(\frac{1}{B_2}\sum_{b_2=1}^{B_2} Y_r\left(\Fhat^{***}_{b_1,b_2}\right)\right)\nonumber\\ 
    &-2\text{Cov}\left(\frac{1}{B_2}\sum_{b_2=1}^{B_2} Y_r\left(\Fhat^{**}_{b_1,b_2}\right),\frac{1}{B_2}\sum_{b_2=1}^{B_2} Y_r\left(\Fhat^{***}_{b_1,b_2}\right)\right)\nonumber\\
    &= \frac{\text{Var}\left( Y_r\left(\Fhat^{**}_{b_1,b_2}\right)\right)}{B_2}+\frac{B_2-1}{B_2}\text{Cov}\left( Y_r\left(\Fhat^{**}_{b_1,b_2}\right), Y_r\left(\Fhat^{**}_{b_1,b'_2}\right)\right)\nonumber\\
    &+ \frac{\text{Var}\left( Y_r\left(\Fhat^{***}_{b_1,b_2}\right)\right)}{B_2}+\frac{B_2-1}{B_2}\text{Cov}\left( Y_r\left(\Fhat^{***}_{b_1,b_2}\right), Y_r\left(\Fhat^{***}_{b_1,b'_2}\right)\right)\nonumber\\
    &-\frac{2}{B_2^2}B_2^2\text{Cov}\left( Y_r\left(\Fhat^{**}_{b_1,b_2}\right), Y_r\left(\Fhat^{***}_{b_1,b_2}\right)\right).\nonumber
\end{align} As proved in \cite{degroot1989}, the variance and covariance between averages of bootstrap samples can be calculated as a function of the variance of the random variable, which for our case is, for a given $r$, $\text{Var}\left( Y_r(\Fhat^*_{b_1})\right)$. Since we are looking at the conditional variance, $\text{Var}\left( Y_r(\Fhat^*_{b_1})\right)$ is no longer random (for brevity we refer to $\text{Var}\left( Y_r(\Fhat^*_{b_1})\right)$ as $\sigma^2$). Therefore, \begin{align}
    \text{Var}\left(\What_r(\Fhat^*_{b_1})\right)&=\frac{\sigma^2(2m^*-1)}{B_2(m^*)^2}\left(1+\frac{1}{B_2}\right)+\frac{(B_2-1)\sigma^2}{B_2m^*}\left(1+\frac{B_2-1}{B_2}\right)-2\frac{\sigma^2}{m^*}. \nonumber
\end{align} Setting the variance of the bias to be in the order of $\Theta\left(\text{Var}\left( Y\left(\Fhat^*_{b_1}\right)\right)/(m^*)^3\right)$ results in\begin{align}
    B_2^*=\Theta\left(\left(\frac{3(m^*)^2-m^*}{m^*+1}\right)^{1/3}\right).\nonumber
\end{align}
\endproof
%\end{proof}
% \section{Examples}
\end{APPENDICES}

% References here (outcomment the appropriate case) 

% CASE 1: BiBTeX used to constantly update the references 
%   (while the paper is being written).
\bibliographystyle{informs2014} % outcomment this and next line in Case 1
\bibliography{myrefs} % if more than one, comma separated

% CASE 2: BiBTeX used to generate mypaper.bbl (to be further fine tuned)
%\input{mypaper.bbl} % outcomment this line in Case 2

\end{document}

%% file: mycommands.tex
\makeatletter
\let\save@mathaccent\mathaccent
\newcommand*\if@single[3]{%
  \setbox0\hbox{${\mathaccent"0362{#1}}^H$}%
  \setbox2\hbox{${\mathaccent"0362{\kern0pt#1}}^H$}%
  \ifdim\ht0=\ht2 #3\else #2\fi
  }
\newcommand*\rel@kern[1]{\kern#1\dimexpr\macc@kerna}
\newcommand*\widebar[1]{\@ifnextchar^{{\wide@bar{#1}{0}}}{\wide@bar{#1}{1}}}
\newcommand*\wide@bar[2]{\if@single{#1}{\wide@bar@{#1}{#2}{1}}{\wide@bar@{#1}{#2}{2}}}
\newcommand*\wide@bar@[3]{%
  \begingroup
  \def\mathaccent##1##2{%
    \let\mathaccent\save@mathaccent
    \if#32 \let\macc@nucleus\first@char \fi
    \setbox\z@\hbox{$\macc@style{\macc@nucleus}_{}$}%
    \setbox\tw@\hbox{$\macc@style{\macc@nucleus}{}_{}$}%
    \dimen@\wd\tw@
    \advance\dimen@-\wd\z@
    \divide\dimen@ 3
    \@tempdima\wd\tw@
    \advance\@tempdima-\scriptspace
    \divide\@tempdima 10
    \advance\dimen@-\@tempdima
    \ifdim\dimen@>\z@ \dimen@0pt\fi
    \rel@kern{0.6}\kern-\dimen@
    \if#31
      \overline{\rel@kern{-0.6}\kern\dimen@\macc@nucleus\rel@kern{0.4}\kern\dimen@}%
      \advance\dimen@0.4\dimexpr\macc@kerna
      \let\final@kern#2%
      \ifdim\dimen@<\z@ \let\final@kern1\fi
      \if\final@kern1 \kern-\dimen@\fi
    \else
      \overline{\rel@kern{-0.6}\kern\dimen@#1}%
    \fi
  }%
  \macc@depth\@ne
  \let\math@bgroup\@empty \let\math@egroup\macc@set@skewchar
  \mathsurround\z@ \frozen@everymath{\mathgroup\macc@group\relax}%
  \macc@set@skewchar\relax
  \let\mathaccentV\macc@nested@a
  \if#31
    \macc@nested@a\relax111{#1}%
  \else
    \def\gobble@till@marker##1\endmarker{}%
    \futurelet\first@char\gobble@till@marker#1\endmarker
    \ifcat\noexpand\first@char A\else
      \def\first@char{}%
    \fi
    \macc@nested@a\relax111{\first@char}%
  \fi
  \endgroup
}
\makeatother
\newcommand{\sumin}{\sum_{i=1}^n}

\newcommand{\oon}{\frac{1}{n}}

\renewcommand*{~}{\relax\ifmmode\sim\else\nobreakspace{}\fi}%GET TILDE FOR "IS DISTRIBUTED AS"

\newcommand{\Ybar}{\bar{Y}}
\newcommand{\Ybbar}{\bar{\Ybar}}

\newcommand{\chat}{\hat{c}}

\newcommand{\Fhat}{\hat{F}}

\newcommand{\What}{\hat{W}}
\newcommand{\betahat}{\hat{\beta}}

\newcommand{\rhohat}{\hat{\rho}}

\newcommand{\thetahat}{\hat{\theta}}

\newcommand{\mcF}{\mathcal{F}}

\newcommand{\mcI}{\mathcal{I}}

\newcommand{\mcN}{\mathcal{N}}
\newcommand{\mcO}{\mathcal{O}}

\newcommand{\mbE}{\mathbb{E}}

\newcommand{\mbI}{\mathbb{I}}

\newcommand{\mbP}{\mathbb{P}}